\documentclass[twoside,journal]{IEEEtran}
    
\usepackage{amsmath,amsfonts}
\usepackage{amssymb}
\usepackage{graphicx}
\usepackage{cite}

\usepackage{algorithmic}
\usepackage{algorithm}

\usepackage{array}
\usepackage{booktabs}
\usepackage{multirow}
\usepackage{stfloats} 
 
\usepackage[caption=false,font=normalsize,labelfont=sf,textfont=sf,labelformat=empty]{subfig}
\usepackage{subcaption} 
\usepackage{textcomp}
\usepackage{url}
\usepackage{verbatim}
\usepackage{pifont}
\usepackage{bm}
\usepackage{xcolor}
\usepackage[pagebackref=true,breaklinks,colorlinks,citecolor=blue,linkcolor=blue,urlcolor=blue,hypertexnames=false]{hyperref}
\newcommand{\RmNum}[1]{\uppercase\expandafter{\romannumeral#1}} 

\begin{document}

\title{UHDRes: Ultra-High-Definition Image Restoration via Dual-Domain Decoupled Spectral Modulation}

\author{Shihao Zhao,~Wanglong Lu\MakeUppercase{*},~Binhao Wang,~Tao Wang,~Kaihao Zhang,~Hanli Zhao\MakeUppercase{*}~
\thanks{S. Zhao, W. Lu,~B. Wang, and~H. Zhao are with the College of Computer Science and Artificial Intelligence, Wenzhou University, Wenzhou, 325000, China.}
\thanks{T. Wang is with the State Key Laboratory for Novel Software Technology, Nanjing University, Nanjing, China.}
\thanks{K. Zhang is with the College of Engineering and Computer Science, Australian National University, Australia.}
\thanks{\MakeUppercase{*}Corresponding authors: W. Lu (email: lwlxhl@gmail.com); H. Zhao (email: hanlizhao@wzu.edu.cn)}%
}

\maketitle

\begin{abstract}
Ultra-high-definition (UHD) images often suffer from severe degradations such as blur, haze, rain, or low-light conditions, which pose significant challenges for image restoration due to their high resolution and computational demands. In this paper, we propose UHDRes, a novel lightweight dual-domain decoupled spectral modulation framework for UHD image restoration. It explicitly models the amplitude spectrum via lightweight spectrum-domain modulation, while restoring phase implicitly through spatial-domain refinement.
We introduce the spatio-spectral fusion mechanism, which first employs a multi-scale context aggregator to extract local and global spatial features, and then performs spectral modulation in a decoupled manner. 
It explicitly enhances amplitude features in the frequency domain while implicitly restoring phase information through spatial refinement.
Additionally, a shared gated feed-forward network is designed to efficiently promote feature interaction through shared-parameter convolutions and adaptive gating mechanisms. 
Extensive experimental comparisons on five public UHD benchmarks demonstrate that our UHDRes achieves the state-of-the-art restoration performance with only 400K parameters, while significantly reducing inference latency and memory usage. The codes and models are available at \url{https://github.com/Zhao0100/UHDRes}.
\end{abstract}


\section{Introduction}
\label{sec:introduction}
Ultra-high-definition (UHD) images have become prevalent across diverse fields, driven by advancements in imaging technology. 
However, UHD images captured under adverse conditions such as low light, fog, or motion blur often experience severe degradation, presenting restoration challenges compounded by high computational requirements. 

\begin{figure}[!t]
    \centering
    \subfloat[]{\includegraphics[width=\columnwidth]{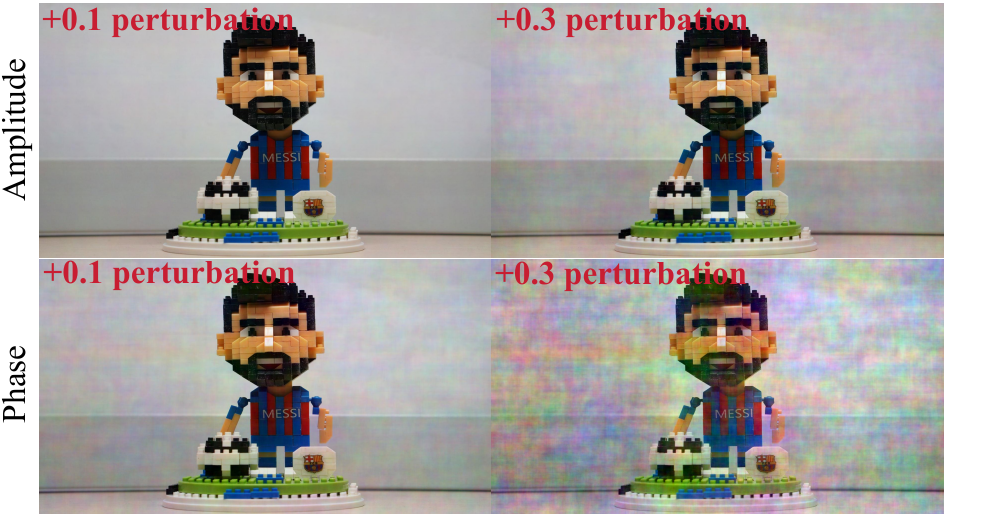}}%
    \vspace{-3em} 
    \subfloat[]{\includegraphics[width=\columnwidth]{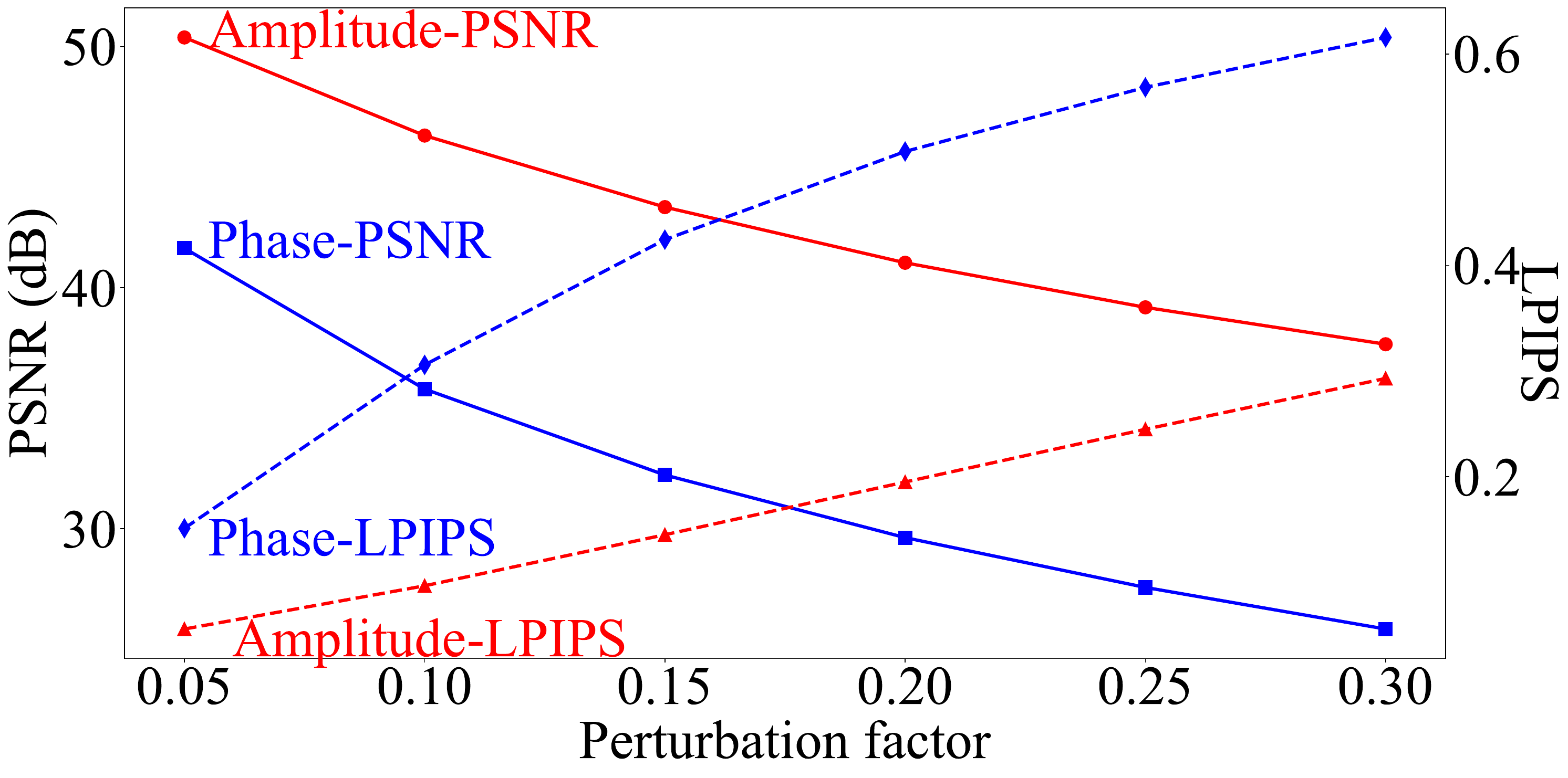}}%
    \vspace{-1.5em}
    \caption{Visualization of the effects of perturbing the amplitude and phase spectra of a given image. (Top) After frequency-domain decomposition, the amplitude and phase spectra are independently perturbed (0.1–0.3) and then inverse-transformed to obtain the visualized reconstructions. (Bottom) PSNR scores of the reconstructed images under different perturbation levels applied to the decomposed amplitude and phase spectra. Under the same degree of direct interference, the phase spectrum is more sensitive.}
    \label{fig:sec2}
\end{figure}

Recent advances~\cite{Uformer,Restormer,TaylorFormer,IFT,SNR-Aware,MambaIR}
have greatly improved image restoration by modeling both local and global dependencies.
However, their complex architectures and heavy computation hinder practical use in high-resolution scenarios.
To address these limitations, recent studies explore frequency-domain modeling, demonstrating that explicitly decomposing images into amplitude and phase spectra better captures diverse degradations~\cite{ FFTformer, FADformer, Wave-Mamba}. 
Specifically, the amplitude spectrum captures global degradation features such as illumination distribution, haze density, or blur intensity, while the phase spectrum precisely encodes the spatial structure details and the spatial arrangement of image components. 
Although these methods achieve impressive results, the simultaneous consideration of both amplitude and phase spectra introduces complexity in the modeling pipeline, making networks more intricate and challenging to train effectively. For example, existing methods may require additional learnable parameters~\cite{UHDFour}, higher GPU memory costs~\cite{AdaIR}, or multiple processing stages (e.g., three-stage frameworks)~\cite{ERR} to ensure satisfactory modeling performance.

As shown in Fig.~\ref{fig:sec2}, we investigate the sensitivity of amplitude and phase components by independently perturbing their values in the frequency domain. While modifying the amplitude introduces relatively smooth changes, even small perturbations (e.g., $+0.1$) in the phase lead to severe degradation in both qualitative appearance and quantitative results. This discrepancy arises from the periodic nature of the phase, which means that small shifts can cause abrupt structural changes, highlighting that the phase is significantly more sensitive and harder to predict accurately in restoration tasks.
This naturally raises an important question: Is it necessary or even beneficial to explicitly model both amplitude and phase simultaneously? 
Can we achieve better results by simplifying the representation learning process?
Specifically, we propose to explicitly model the robust amplitude component while implicitly capturing the sensitive phase information in the spatial domain.

\begin{figure}[!t]
    \centering
    \includegraphics[width=\columnwidth]{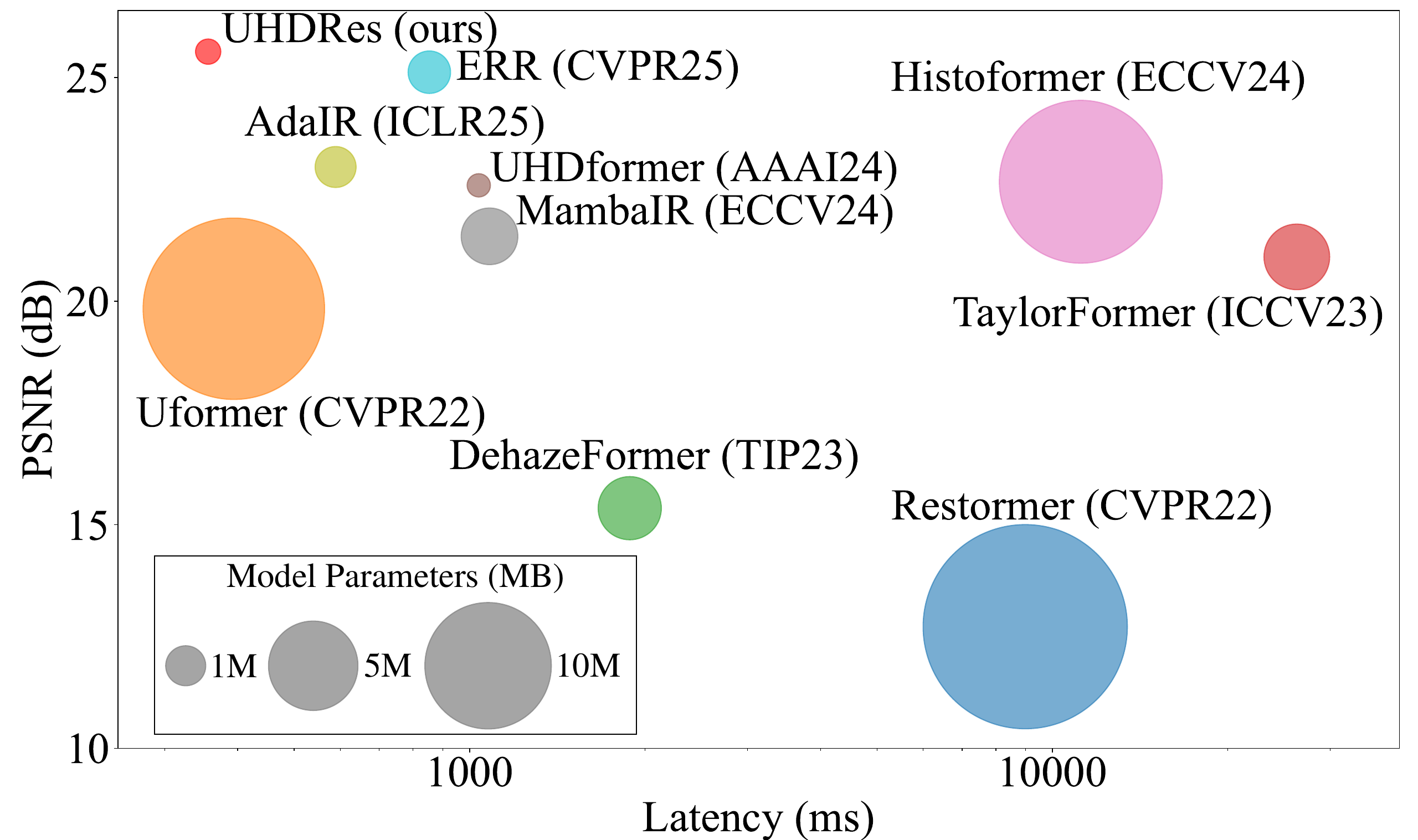} 
    \caption{Model performance and computational efficiency comparison on $1024 \times 1024$ resolution images from the UHD-Haze dataset. PSNR, inference latency, and parameter count are visualized. UHDRes achieves the best trade-off between restoration quality, inference speed, and model size.}
    \label{fig:intro}
\end{figure}

To tackle the complex degradations in UHD image restoration, we explore the use of a dual-domain strategy that leverages the complementary strengths of spatial and frequency representations. The spatial domain is well-suited for capturing fine-grained textures and structural layouts, while the frequency domain provides a compact and global view of image degradations. However, directly optimizing both the amplitude and phase spectra in the frequency domain is problematic, as the phase is highly sensitive and unstable. 
Motivated by this observation, we propose to explicitly model and modulate the amplitude spectrum to capture robust global degradation patterns, while adopting an identity mapping for phase information, which is implicitly restored through spatial-domain processing. 
This design reduces modeling complexity, improves training stability, and achieves a strong balance between restoration performance and computational efficiency.

To this end, we propose UHDRes, a dual-domain decoupled spectral modulation framework that enables high-quality and efficient UHD image restoration by integrating two complementary mechanisms.
We first introduce the spatio-spectral fusion mechanism that leverages multi-scale large-kernel convolutions to capture both local textures and global structures. 
These global-local representations are then projected into the frequency domain.
To address global degradations, we propose a decoupled spectral modulation that explicitly modulates the amplitude spectrum while restoring the phase spectrum implicitly through subsequent structural refinement. 
Secondly, a shared gated feed-forward network is designed to improve representational capacity and efficiency.
It adopts a dual-branch architecture utilizing pointwise convolution and depthwise convolution to achieve efficient feature interaction with minimal computational overhead.
Parameter sharing across branches enforces structural consistency and promotes general feature learning, while a lightweight gating strategy enables adaptive filtering and implicit disentanglement of features.
Leveraging the proposed components, UHDRes is designed to robustly address a wide range of degradation scenarios, achieving both high restoration quality and computational efficiency, as demonstrated in Fig.~\ref{fig:intro}.

Our main contributions are summarized as follows:
\begin{itemize}
\item We propose UHDRes, a novel lightweight framework for UHD image restoration, which features a dual-domain decoupled spectral modulation design that jointly exploits spatial-domain structures and frequency-domain features in a decoupled manner.
\item We introduce a novel spatio-spectral fusion mechanism that explicitly enhances amplitude features in the frequency domain while implicitly restoring phase information through spatial refinement.
\item We design a novel shared gated feed-forward network that leverages lightweight convolutions, weight sharing, and adaptive gating to further enhance the restored feature representations.
\item We conducted extensive experimental comparisons on five public UHD benchmarks (UHD-LL, UHD-Haze, 8KDehaze-mini, UHD-Blur, and 4K-Rain13k), demonstrating that our UHDRes achieves the state-of-the-art restoration performance with only 400K parameters.
\end{itemize}

\section{Related Work}
\label{sec: related work}
\begin{figure*}[t]
    \centering
    \includegraphics[width=\textwidth]{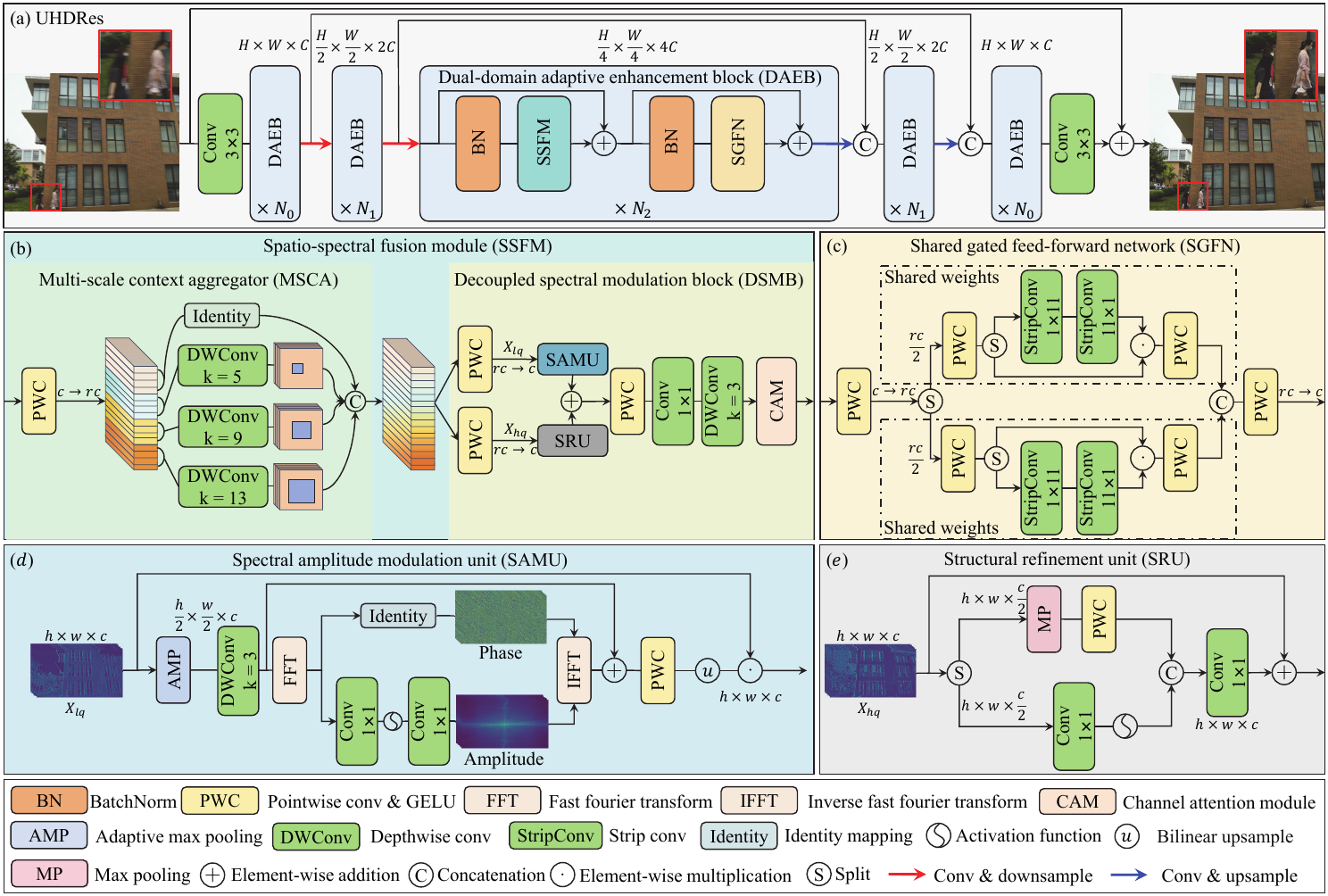} 
    \caption{Overall network architecture of UHDRes. The model consists of stacked dual-domain adaptive enhancement blocks (DAEBs), each comprising a spatio-spectral fusion module (SSFM) and a shared gated feed-forward network (SGFN). SSFM combines multi-scale feature extraction with frequency-domain decoupled modulation. The multi-scale context aggregator (MSCA) captures local-global textures, while the decoupled spectral modulation block (DSMB) adjusts amplitude via the spectral amplitude modulation unit (SAMU) and refines structure using the structural refinement unit (SRU). SGFN promotes efficient and consistent feature learning through shared dual-branch gating.
    } 
    \label{fig:Model}
\end{figure*}

Image restoration aims to recover high-quality visual content from degraded inputs, covering sub-tasks such as low-light enhancement~\cite{IFT, SNR-Aware, DiffLL}, dehazing~\cite{DehazeFormer, TaylorFormer,WANG2024109956}, deblurring~\cite{MIMO-Unet++, Stripformer, FFTformer, Restormer, Uformer, MambaIR, AdaIR,LU2025111312}, and deraining~\cite{JORDER-E, RCDNet, SPDNet, IDT, DRSformer, UDR-Mixer}. 
Previous methods were typically designed for low-resolution images, and the high-resolution characteristics of UHD images pose greater challenges for restoration, as this requires a better balance between performance and efficiency~\cite{Binhao_FST}. 

Recent works have then focused on solving the restoration of UHD images using either task-specific models for particular degradations~\cite{UHD, LLFormer, UHDFour, UDR-Mixer, Wave-Mamba} or general-purpose architectures~\cite{MixNet, UHDformer, TriFormer, D2Net, ERR}.
Nevertheless, they still have room to balance performance and efficiency adequately.
Several methods have been proposed to mitigate this issue~\cite{UHDFour,UHDformer,Wave-Mamba,ERR}. UHDFour~\cite{UHDFour} and UHDformer~\cite{UHDformer} adopt high downsampling ratios to alleviate the computational burden imposed by UHD images. Even though these methods leverage information from the original resolution space for guidance, such approaches inevitably leads to the loss of rich detail and structural information inherent in UHD images, thereby limiting performance in cases with complex degradation information. 
ERR~\cite{ERR} introduced a dedicated architecture for UHD restoration using progressive frequency decoupling, achieving high fidelity results. However, its sequential three-level design increases training time, and its reliance solely on frequency-domain cues limits performance in blurry scenes. 
In contrast, our UHDRes employs a hybrid restoration strategy that integrates frequency and spatial domains to simultaneously capture global degradations and fine-grained texture details, enhancing computational efficiency for UHD image restoration while maintaining high performance.

To achieve a better trade-off between performance and efficiency in UHD image restoration, frequency-domain modeling has recently emerged as a promising direction.
Processing images in the frequency domain can better facilitate global information interaction and decouple the intrinsic properties of images~\cite{UHDFour,DFSMN,FADformer,FDTANet,FFTformer, ERR}.
UHDFour~\cite{UHDFour} found that frequency domain amplitude information is related to low-light conditions and attempted to decouple frequency domain features in polar coordinates for direct separate modeling, thereby eliminating darkness and noise. FADformer~\cite{FADformer} discovered that rain streak patterns are correlated with frequency domain information, obtaining complex domain features via Fast Fourier Transform and modeling the real and imaginary parts separately. AdaIR~\cite{AdaIR}, on the other hand, achieves high-low frequency separation by adaptively masking the frequency domain features of the input image and injecting intermediate features to learn degradation patterns. 
However, these methods often overlook the sensitivity of phase information, which is highly correlated with edge and textural details, making it difficult to balance computational efficiency and performance when applied to UHD image restoration.
Our dual-domain decoupled spectral modulation framework explicitly enhances amplitude features in the frequency domain while restoring phase implicitly via subsequent spatial refinement to enhance the feature representation capability.

\section{Method}
\label{sec:method}
\subsection{Overall pipeline (UHDRes)}
As shown in Fig.~\ref{fig:Model} (a), we propose {UHDRes}, a dual-domain decoupled spectral modulation framework for UHD image restoration. 
To effectively utilize both spatial and frequency information, we first design the {spatio-spectral fusion module (SSFM)}. 
It consists of a multi-scale context aggregator (MSCA) and a decoupled spectral modulation block (DSMB).
The MSCA  uses large-kernel convolutions at multiple scales to extract rich local and global features in the spatial domain.  
To better capture the degradations that are more prominent in the frequency domain, the DSMB first transforms features into the frequency domain and applies the {spectral amplitude modulation unit (SAMU)} to explicitly adjust the amplitude spectrum. 
Meanwhile, the phase is adjusted via identity mapping and further refined by the {structural refinement unit (SRU)}. 
This design enables the network to capture global degradation patterns in the spectral domain while restoring fine structural details in the spatial domain.
We further introduce the shared gated feed-forward network (SGFN). By sharing weights across its dual-gated branches, SGFN guides the network to learn a unified representation, thereby ensuring structural consistency. 
Our framework allows the network to simplify representation learning, reduce redundancy, and maintain high-fidelity structural recovery.

Given a degraded input image $I_{LQ} \in \mathbb{R}^{H \times W \times 3}$ (with height $H$, width $W$ and, 3 channels), our UHDRes predicts a residual map $R \in \mathbb{R}^{H \times W \times 3}$, which is added to the input image to obtain the restored image, i.e., $I_{HQ} = I_{LQ} + R$.
UHDRes first applies a $3 \times 3$ convolution to extract shallow features $I \in \mathbb{R}^{H \times W \times C}$, with $C$ as the initial channel number. 
These features are then processed by a 3-level multi-scale encoder-decoder architecture, where each scale consists of multiple dual-domain adaptive enhancement blocks (DAEBs).
Each DAEB integrates SSFM and SGFN, along with batch normalization and residual connections. 

\subsection{Dual-domain adaptive enhancement block (DAEB)}
The DAEB is the core building block of UHDRes, which effectively aggregates dual-domain information by adaptively enhancing feature representation capabilities in both spatial and frequency domains. As shown in Fig.~\ref{fig:Model} (a), given the input feature map $X_{in} \in \mathbb{R}^{h \times w \times c}$ (with height $h$, width $w$, and $c$ channels), the entire process can be expressed as:
\begin{equation}
\begin{aligned}
& X_{SSFM} = X_{in} + \operatorname{SSFM}(\operatorname{BN}(X_{in})), \\
& X_{out} = X_{SSFM} + \operatorname{SGFN}(\operatorname{BN}(X_{SSFM})),
\end{aligned}
\end{equation}
where $\operatorname{SSFM}(\cdot)$ denotes the spatio-spectral fusion module, $\operatorname{SGFN}(\cdot)$ represents the shared gated feed-forward network, and $\operatorname{BN}(\cdot)$ refers to the Batch Normalization.

\subsection{Spatio-spectral fusion module (SSFM)}
As shown in Fig.~\ref{fig:Model} (b), our spatio-spectral fusion module (SSFM) integrates the multi-scale context aggregator (MSCA) and decoupled spectral modulation block (DSMB) for hybrid modulation in both spatial and frequency domains.
Within the MSCA, it effectively extracts multi-scale feature information using parallel multi-scale large-kernel convolutions. 
In the DSMB, we adopt a dual-branch structure to decouple and modulate the amplitude and phase spectra, enabling the network to efficiently learn according to the distinct characteristics of different frequency domain signals. This approach allows for global modeling while maintaining lower computational overhead. 

\textbf{Multi-scale context aggregator (MSCA).} 
In the multi-scale context aggregator, we extract and enhance feature information across different frequency bands using parallel multi-scale large-kernel convolutions. Given the input feature $X \in \mathbb{R}^{h \times w \times c}$, our MSCA can be expressed as:
\begin{equation}
\begin{aligned}
& X_0, X_1, X_2, X_3, = \operatorname{Split}_4(\operatorname{PWC}(X)),  \\
& X_{0}{'}=\operatorname{Identity}(X_{0}),  \\
& X_{m}{'} = \mathcal{F}_{k_m}(X_{m}), \quad m \in \{1, 2, 3\},  \\
& X_{MSCA} = \operatorname{Concat}(X_{0}{'},X_{1}{'},X_{2}{'},X_{3}{'}),  \\
\end{aligned}
\end{equation}
where $\operatorname{PWC}(\cdot)$ denotes pointwise convolution (i.e., $1 \times 1$ convolution). The function $\operatorname{Split}_4(\cdot)$ evenly divides the input along the channel dimension to obtain four feature groups. $\operatorname{Identity}(\cdot)$ represents the identity mapping operation. $\mathcal{F}_{k_m}(\cdot)$ denotes the operation of the $m$-th convolutional branch, which represents a depthwise convolution layer with a kernel size of $k_m$, and $k_m \in \{5, 9, 13\}$. $\operatorname{Concat}(\cdot)$ denotes concatenation along the channel dimension.
Our MSCA first undergoes pointwise convolution to expand its channel dimension from $c$ to $rc$, where $r=2$. Then, it is split along the channel dimension into $X_{i}, i \in \{0, 1, 2, 3\}$. Among them, $X_{0}$ is adjusted through identity mapping, while the others are processed by convolutions with varying receptive fields.
After channel concatenation, $X_{MSCA}$ is obtained for the subsequent frequency amplitude adaptive modeling.

\textbf{Decoupled spectral modulation block (DSMB).} 
In the decoupled spectral modulation block, frequency domain signals are decoupled and modeled through a dual-branch structure. Specifically, global structure and degradation information are modulated in the spectral amplitude modulation unit (SAMU), while fine-grained texture details are complemented in the structural refinement unit (SRU). 
Within the DSMB, low-frequency features $X_{lf}$ and high-frequency features $X_{hf}$ are first adaptively extracted from $X_{MSCA}$, which contains rich multi-scale information. These features are then fed into the SAMU and SRU, respectively. They are further fused and enhanced through a Channel Attention Module (CAM) to achieve comprehensive frequency domain modeling. 
The process is as follows:
\begin{equation}
\begin{aligned}
& X_{lf} = \operatorname{PWC}(X_{MSCA}),X_{hf} = \operatorname{PWC}(X_{MSCA}),  \\
& X_{fre} = \operatorname{PWC}(\operatorname{SAMU}(X_{lf})+\operatorname{SRU}(X_{hf})),  \\
& X_{DSMB} = \operatorname{CAM}(\operatorname{DWConv_{3 \times 3}}(\operatorname{Conv_{1 \times 1}}(X_{fre}))),
\end{aligned}
\end{equation}
where $\operatorname{SAMU}(\cdot)$ denotes the spectral amplitude modulation unit, $\operatorname{SRU}(\cdot)$ denotes the structural refinement unit, $\operatorname{Conv_{3 \times 3}}(\cdot)$ denote a standard convolution with the kernel size of $3 \times 3$, $\operatorname{DWConv_{3 \times 3}(\cdot)}$ is a depthwise convolution with the kernel size of $3 \times 3$, and $\operatorname{CAM}(\cdot)$ refers to the channel attention module~\cite{CAM}.

\textbf{Spectral amplitude modulation unit (SAMU).} 
As shown in Fig~\ref{fig:Model} (d), we primarily focus on restoring the image's low-frequency structures and degradation information. By explicitly transforming features into the frequency domain via FFT, and based on observations such as those in Fig.~\ref{fig:sec2}, we selectively model only the amplitude spectrum while maintaining an identity mapping for the phase component. This allows the model in SAMU to concentrate on recovering global degradation information related to the amplitude spectrum. 

Specifically, we first apply max pooling to the low-frequency features $X_{lf}$, followed by a depthwise convolution for feature enhancement. Subsequently, FFT is employed to extract the amplitude spectrum $A_{lf}$ and phase spectrum $P_{lf}$ of the feature. Here, modulation is applied to the amplitude spectrum $A_{lf}$, while identity mapping is adopted for the phase spectrum $P_{lf}$, thereby yielding the modulated amplitude $A'_{lf}$ and phase $P'_{lf}$, as shown below:
\begin{equation}
\begin{aligned}
& X_{lf\_dw} = \operatorname{DWConv_{3 \times 3}}(\operatorname{AMP}(X_{lf})),  \\
& Z = \mathcal{F}_{FFT}(X_{lf\_dw}), A_{lf} = |Z|,P_{lf}=\operatorname{arg}(Z),  \\
& A'_{lf} = \operatorname{Conv_{1 \times 1}}(\operatorname{LeakyReLU}(\operatorname{Conv_{1 \times 1}}(A_{lf}))),  \\
& P'_{lf} = \operatorname{Identity}(P_{lf}),  \\
\end{aligned}
\end{equation}
where $\operatorname{AMP}(\cdot)$ denotes an adaptive max pooling operation that halves spatial dimensions, $\mathcal{F}_{FFT}(\cdot)$ denotes the Fast Fourier Transform, $|\cdot|$ denotes the modulus calculation, $\arg(\cdot)$ denotes the argument calculation of a complex number, and $\operatorname{LeakyReLU}(\cdot)$ represents the activation function with a negative slope set to 0.1.

Subsequently, the complex spectrum $Z'$ is reconstructed from its amplitude $A'_{lf}$ and phase $P'_{lf}$ using polar representation based on Euler's formula.
The frequency-domain signal is transformed back into the spatial domain via IFFT and connected with the previous features through a residual connection to maintain the integrity of the feature structure. Then, it is upsampled back to its original size, and $X_{lf}$ is enhanced through a gated mechanism. 
It is expressed as:
\begin{equation}
\begin{aligned}
& Z' = A'_{lf} \cdot e^{j \cdot P'_{lf}}, \label{eq:faam_complex_reconstruct} \\
& X_{att} = \operatorname{PWC}(\mathcal{F}_{IFFT}(Z')+X_{lf\_dw}),  \\
& \operatorname{SAMU}(X_{lf}) = \operatorname{UP}(X_{att}) \odot X_{lf},  \\
\end{aligned}
\end{equation}
where $j$ is the imaginary unit, $\mathcal{F}_{IFFT}(\cdot)$ denotes the Inverse Fast Fourier Transform~\cite{Fourllie,UHDFour}, $\operatorname{UP}(\cdot)$ signifies the bilinear upsampling operation used to restore the original spatial resolution, and $\odot$ signifies the element-wise product operation.

\textbf{Structural refinement unit (SRU).} 
In recent methods applied to UHD image restoration~\cite{UHDformer,D2Net,ERR}, the importance of high-frequency information is often overlooked, leading to suboptimal restoration performance in scenarios with image blur. As shown in Fig~\ref{fig:Model} (e), we introduce the structural refinement unit (SRU), which focuses on restoring fine-grained texture details. This implicitly complements the phase spectrum, thereby meeting the demands of various degradation scenarios.

Specifically, the high-frequency features $X_{hf}$ is first split along the channel dimension into $X_{hf1}$ and $X_{hf2}$. We enrich texture details and supplement crucial high-frequency information by performing local operations on the high-frequency components $X_{hf1}$ and $X_{hf2}$. Subsequently, these high-frequency components are concatenated along the channel dimension and further fused through a standard convolution. Then, a residual connection with $X_{hf}$ is applied to maintain semantic integrity. The entire process is as follows:
\begin{equation}
\begin{aligned}
& X_{hf1},X_{hf2} = \operatorname{Split}_2(X_{hf}),  \\
& {X'_{hf1}} = \operatorname{PWC}(\operatorname{MP}(X_{hf1})),  \\
& {X'_{hf2}} = \operatorname{GELU}(\operatorname{Conv_{3 \times 3}}(X_{hf2})),  \\
& \operatorname{SRU}(X_{hf}) = \operatorname{Conv_{3 \times 3}}(\operatorname{Concat}({X'_{hf1}},{X'_{hf2}}))+X_{hf},  \\
\end{aligned}
\end{equation}
where $\operatorname{MP}(\cdot)$ denotes a max pooling operation with kernel size $3$ and stride $1$,
the function $\operatorname{Split}_2(\cdot)$ evenly divides the input along the channel dimension to obtain two feature groups, $\operatorname{Conv_{3 \times 3}}(\cdot)$ denote standard convolutions with the kernel size of $3 \times 3$, and $\operatorname{GELU}(\cdot)$ stands for Gaussian error linear unit activation function. 

\subsection{Shared gated feed-forward network (SGFN)}
After hybrid-domain enhancement, it is necessary to further enhance the restored feature representations from different domains. To this end, we design a novel shared gated feed-forward network (SGFN). SGFN introduces a dual-gating mechanism employing horizontal and vertical strip convolutions to further fuse information from both domains, and by utilizing shared weights, it enhances the feature representation capability.

As shown in Fig.~\ref{fig:Model} (c), given an input feature map $X' \in \mathbb{R}^{h \times w \times c}$, $X'$ first undergoes channel expansion and is then divided along the channel dimension into two tensors $Z_0, Z_1 \in \mathbb{R}^{h \times w \times c}$. $Z_0$ and $Z_1$ are processed by two parallel branches that share the same set of weights, as follows:
\begin{equation}
\begin{aligned}
& Z_0, Z_1 = \operatorname{Split}_2(\operatorname{PWC}(X')),  \\
& Q_i,V_i = \operatorname{Split}_2(\operatorname{PWC}(Z_i)), \quad i \in \{0,1\}, 
\end{aligned}
\end{equation}
where $Q_0$ and $Q_1$ represent the attention maps, while $V_0$ and $V_1$ serve as the value tensors.
Then, we enhance the feature representation by using lightweight convolutions, weight sharing, and adaptive gating, as expressed below:
\begin{equation}
\begin{aligned}
& Q'_i = \operatorname{DWConv_{11 \times 1}}(\operatorname{DWConv_{1 \times 11}}(Q_i)), \quad i \in \{0,1\},  \\
& Z'_i = \operatorname{PWC}(V_i \odot Q'_i), \quad i \in \{0,1\},  \\
& \operatorname{SGFN}(X') = \operatorname{PWC}(\operatorname{Concat}(Z'_0, Z'_1)),
\end{aligned}
\end{equation}
where $\operatorname{DWConv_{1 \times 11}(\cdot)}$ and $\operatorname{DWConv_{11 \times 1}}(\cdot)$ represent depthwise convolutions with kernel sizes of $1 \times 11$ and $11 \times 1$, respectively, which are used to capture horizontal and vertical spatial information. 
$Q_0$ and $Q_1$ are simultaneously and sequentially processed by depthwise separable convolutions with kernel sizes of $1 \times 11$ and $11 \times 1$ respectively, sharing the same weights, yielding their respective gated attention maps $Q'_0$ and $Q'_1$. Each branch output is obtained by element-wise multiplying its feature map with the corresponding value tensor. Then, these outputs are concatenated along the channel dimension, followed by element-wise convolution to further enhance features, thereby obtaining the output.

\subsection{Loss functions}
We train our model in both the spatial and frequency domains by minimizing the total loss, which is expressed as:
\begin{equation}
\mathcal{L}_{total} = \mathcal{L}_{pixel} + \lambda \cdot \mathcal{L}_{freq}, \label{eq:total_loss}
\end{equation}
where $\mathcal{L}_{pixel}$ represents the pixel-wise loss function, $\mathcal{L}_{freq}$ is the frequency loss, and $\lambda$ is the loss weight to balance the two loss functions.

The $\mathcal{L}_{pixel}$ is defined as: $\mathcal{L}_{pixel} = \|I_{HQ} - I_{GT}\|_1$,
where $\left\| \cdot \right\|_1$ denotes the $L_{1}$ norm, and $I_{HQ}$ and $I_{GT}$ represent the restored high-quality result and the corresponding ground-truth image.
The $\mathcal{L}_{freq}$ is expressed as: $\mathcal{L}_{freq} = \|\mathcal{F}_{FFT}(I_{HQ}) - \mathcal{F}_{FFT}(I_{GT})\|_1$. In this paper, we empirically set $\lambda = 0.1$ following the established practice~\cite{seemore,SAFMN}.

\section{Experiments}
\label{sec:experiments}
In this section, we evaluated the performance of our UHDRes by comparing with the state-of-the-art (SOTA) algorithms on four standard UHD image restoration tasks at various resolutions: low-light image enhancement (4K), image dehazing (4K and 8K), image deblurring (4K), and image deraining (4K).
\subsection{Experimental settings}

\textbf{Datasets.} 
For the low-light image enhancement task, we employed the UHD-LL dataset~\cite{UHDFour}, which comprises 2,150 pairs of 4K images (2,000 for training and 150 for testing). For image dehazing, we leveraged the UHD-Haze dataset~\cite{UHDformer} and 8KDehaze-mini~\cite{dehazexl}. UHD-Haze dataset provides 2,521 pairs of 4K images, with 2,290 pairs designated for training and 231 for testing. The 8KDehaze-mini dataset includes 100 images of $8192 \times 8192$ (8K) resolution. For the image deblurring task, we utilized the UHD-Blur dataset~\cite{UHDformer}, consisting of 2,264 pairs of 4K images (1,964 for training and 300 for testing). Lastly, for the image deraining task, we utilized the 4K-Rain13k dataset~\cite{UDR-Mixer}, which contains 13,000 pairs of 4K images (12,500 for training and 500 for testing).

\textbf{Compared methods.} 
For the low-light image enhancement task, we compared our method with several current the SOTA approaches on the UHD-LL dataset including IFT~\cite{IFT}, SNR-Aware~\cite{SNR-Aware}, LLFormer~\cite{LLFormer}, DiffLL~\cite{DiffLL}, UHDFour~\cite{UHDFour}, LMAR~\cite{LMAR}, Wave-Mamba~\cite{Wave-Mamba}, Restormer~\cite{Restormer}, Uformer~\cite{Uformer}, UHDformer~\cite{UHDformer}, MambaIR~\cite{MambaIR}, AdaIR~\cite{AdaIR}, UHDDIP~\cite{UHDDIP} and ERR~\cite{ERR}. The quantitative results of LMAR were taken from D\textsuperscript{2}R-UHDNet~\cite{CIMF-Net}. 
We re-trained MambaIR and AdaIR to obtain their quantitative and qualitative results, while the quantitative results for other comparative methods were obtained from ERR~\cite{ERR}.

For the dehazing task, we compared UHDRes with the recent SOTA methods on the UHD-Haze dataset, including UHD~\cite{UHD}, DehazeFormer~\cite{DehazeFormer}, TalorFormer~\cite{TaylorFormer}, Histoformer~\cite{Histoformer}, Restormer~\cite{Restormer}, Uformer~\cite{Uformer}, UHDformer~\cite{UHDformer}, MambaIR~\cite{MambaIR}, AdaIR~\cite{AdaIR}, D2Net~\cite{D2Net}, UHDDIP~\cite{UHDDIP}, and ERR~\cite{ERR}. The quantitative results of TalorFormer were taken from D\textsuperscript{2}R-UHDNet~\cite{CIMF-Net}, and those for D2Net were taken from its original paper. 
We re-trained Histoformer, MambaIR, and AdaIR to obtain their quantitative and qualitative results, while the qualitative results for other comparative methods were obtained from ERR~\cite{ERR}. 
Additionally, we retrained several SOTA methods on the 8KDehaze-mini dataset~\cite{dehazexl} with higher resolutions to compare with our approach, including: Restormer~\cite{Restormer}, Histoformer~\cite{Histoformer}, UHDformer~\cite{UHDformer}, AdaIR~\cite{AdaIR}, and ERR~\cite{ERR}.

\begin{figure*}[t]
    \centering
    \includegraphics[width=\textwidth]{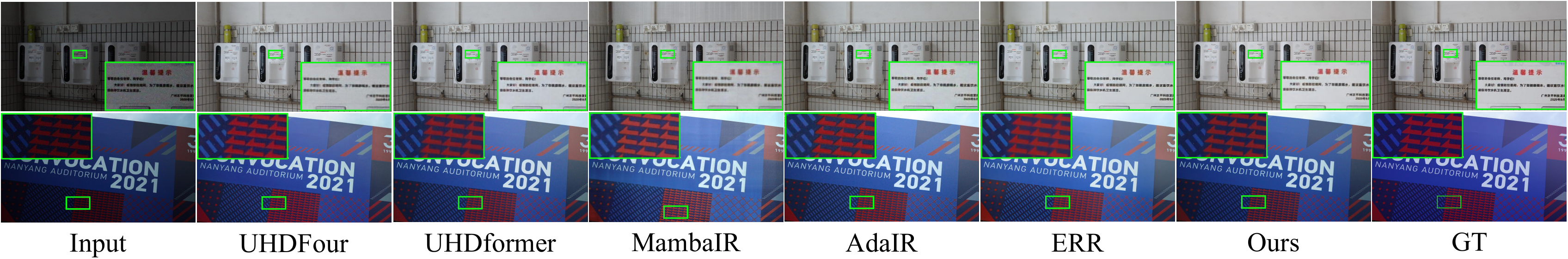} 
    \caption{Visual comparison to SOTA methods on the UHD-LL dataset. UHDRes restores more realistic colors.}
    \label{fig:UHDLL}
\end{figure*}

\begin{figure*}[t]
    \centering
    \includegraphics[width=\textwidth]{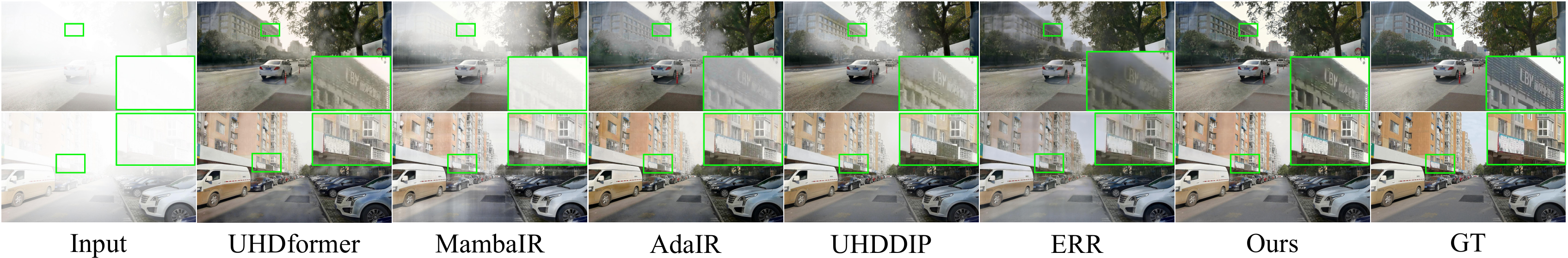}
    \caption{Visual comparison to SOTA methods on the UHD-Haze dataset. UHDRes removes thick fog and restores richer details.}
    \label{fig:UHDHaze}
\end{figure*}

For the deblurring task, we compared with the recent SOTA methods on the UHD-Blur dataset, including MIMO-Unet++~\cite{MIMO-Unet++}, Stripformer~\cite{Stripformer}, FFTformer~\cite{FFTformer}, MixNet~\cite{MixNet}, TriFormer~\cite{TriFormer}, Restormer~\cite{Restormer}, Uformer~\cite{Uformer}, UHDformer~\cite{UHDformer}, MambaIR~\cite{MambaIR}, AdaIR~\cite{AdaIR}, D2Net~\cite{D2Net}, UHDDIP~\cite{UHDDIP} and ERR~\cite{ERR}. The quantitative results of MixNet, TriFormer, and D2Net were sourced from their respective original papers.
We re-trained MambaIR and AdaIR to obtain their quantitative and qualitative results, while the quantitative results of other comparative methods were obtained from ERR~\cite{ERR}.

For the deraining task, we conducted comparisons with the SOTA methods on the 4K-Rain13k dataset, including JORDER-E~\cite{JORDER-E}, RCDNet~\cite{RCDNet}, SPDNet~\cite{SPDNet}, IDT~\cite{IDT}, DRSformer~\cite{DRSformer}, UDR-S2Former~\cite{DRSformer}, UDR-Mixer~\cite{UDR-Mixer}, Restormer~\cite{Restormer} and ERR~\cite{ERR}. The quantitative and qualitative results of ERR~\cite{ERR} are taken directly from its original paper, while those of other comparative methods are sourced from UDR-Mixer~\cite{UDR-Mixer}.

\textbf{Implementation.} 
Our model employs an end-to-end approach and applies uniform parameter settings across four distinct restoration tasks. Within UHDRes, the number of DAEB differs per level, as indicated by $N_0, N_1,$ and $N_2$ in Fig.~\ref{fig:Model}, which correspond to $[2, 3, 4]$, respectively. The initial channel number is $C=12$, and from level 1 to level 3, the channel numbers are $[12, 24, 48]$ respectively. Furthermore, the expansion factor $r$ within each DAEB is 2.

Histoformer~\cite{Histoformer}, MambaIR~\cite{MambaIR}, and AdaIR~\cite{AdaIR} were re-trained based on their officially released codes.
To better suit the 4K image restoration task, we halved the channel numbers and the number of modules per layer for MambaIR and AdaIR, reducing their parameter counts to 2.01M and 1.05M, respectively. 
As Histoformer and MambaIR cannot process full-resolution images within 24GB of GPU memory, we followed the established practice~\cite{UHDFour,UHDformer} by performing inference on downsampled inputs, which are 0.5$\times$ for Histoformer and 0.6$\times$ for MambaIR.

Our UHDRes and the re-trained models (Histoformer~\cite{Histoformer}, MambaIR~\cite{MambaIR}, and AdaIR~\cite{AdaIR}) were trained using the same settings. All methods were optimized using the AdamW optimizer with parameters $\beta_1=0.9$ and $\beta_2=0.999$. 
The initial learning rate was uniformly set to $5 \times 10^{-4}$.
Specifically, for the UHD-LL, UHD-Haze, UHD-Blur, and 4K-Rain13k datasets, a patch size of 512 was utilized, and the total batch size was set to 12. 
A cosine annealing strategy was then employed to reduce the learning rate from its initial value to $1 \times 10^{-7}$ over $1 \times 10^{5}$ iterations.
For the 8KDehaze-mini dataset, images were center-cropped to a size of $4320 \times 7860$ to match the aspect ratios commonly found in industrial applications. During training, a patch size of 1024$^2$ and a total batch size of 2 were used. Similarly, a cosine annealing strategy was applied to reduce the learning rate from its initial value to $1 \times 10^{-6}$ over $1 \times 10^{4}$ iterations.
All experiments were conducted on four NVIDIA 4090 GPUs.

\textbf{Evaluation.}    
Following prior research, the peak signal to noise ratio (PSNR)~\cite{PSNR} and the structural similarity (SSIM)~\cite{SSIM} were utilized for evaluating model performance, and the learned perceptual image patch similarity (LPIPS)~\cite{LPIPS} was employed to assess perceptual performance. Beyond quantitative results, we also compared models based on their parameter count, inference latency, and maximum memory usage during inference. 

\begin{table}[t]
\caption{Comparison of quantitative results on UHD-LL. UHDRes achieves the highest PSNR with a relatively small number of parameters.\label{tab:UHDLL}}
\setlength{\tabcolsep}{2pt}
\resizebox{\columnwidth}{!}{%
\centering
\begin{tabular}{cccccc}
\hline
\multirow{2}{*}{Method}      & \multirow{2}{*}{Source} & \multirow{2}{*}{Params$\downarrow$} & \multirow{2}{*}{PSNR$\uparrow$}   & \multirow{2}{*}{SSIM$\uparrow$}   & \multirow{2}{*}{LPIPS$\downarrow$} \\
                             &                         &                                     &                                   &                                   &                                    \\ \hline
IFT~\cite{IFT}               & ICCV'21                 & 11.56M                              & 21.960                            & 0.8700                            & 0.324                              \\
SNR-Aware~\cite{SNR-Aware}   & CVPR'22                 & 40.08M                              & 22.720                            & 0.8770                            & 0.304                              \\
Restormer~\cite{Restormer}   & CVPR'22                 & 26.1M                               & 21.536                            & 0.8437                            & 0.361                              \\
Uformer~\cite{Uformer}       & CVPR'22                 & 20.6M                               & 21.303                            & 0.8233                            & -                                  \\
LLFormer~\cite{LLFormer}     & AAAI'23                 & 13.2M                               & 22.790                            & 0.8530                            & 0.264                              \\
DiffLL~\cite{DiffLL}         & TOG'23                  & 17.29M                              & 21.360                            & 0.8720                            & 0.239                              \\
UHDFour~\cite{UHDFour}       & ICLR'23                 & 17.5M                               & 26.226                            & 0.9000                            & 0.239                              \\
UHDformer~\cite{UHDformer}   & AAAI'24                 & \textcolor{red}{\textbf{0.34M}}     & 27.113                            & 0.9271                            & 0.245                              \\
LMAR~\cite{LMAR}             & CVPR'24                 & 1.97M                               & 26.270                            & 0.9196                            & 0.225                              \\
Wave-Mamba~\cite{Wave-Mamba} & ACM MM'24               & 1.26M                               & 27.350                            & 0.9130                            & \textcolor{red}{\textbf{0.185}}    \\
MambaIR~\cite{MambaIR}       & ECCV'24                 & 2.01M                               & 24.127                            & 0.8965                            & 0.284                              \\
AdaIR~\cite{AdaIR}           & ICLR'25                 & 1.05M                               & 25.669                            & 0.9014                            & 0.268                              \\
UHDDIP~\cite{UHDDIP}         & TCSVT'25                & 0.81M                               & 26.749                            & 0.9281                            & \textcolor{blue}{\textbf{0.208}}   \\
ERR~\cite{ERR}               & CVPR'25                 & 1.13M                               & \textcolor{blue}{\textbf{27.570}} & \textcolor{red}{\textbf{0.9320}}  & 0.214                              \\
UHDRes                       & Ours                    & \textcolor{blue}{\textbf{0.40M}}    & \textcolor{red}{\textbf{27.693}}  & \textcolor{blue}{\textbf{0.9311}} & 0.231                              \\ \hline
\end{tabular}

}
\end{table}

\subsection{Comparisons with SOTA methods}

\textbf{UHD low-light image enhancement results.} 
Table~\ref{tab:UHDLL} presents the quantitative performance of the compared methods on the UHD-LL dataset. 
In general, most existing methods have demonstrated strong performance on low-light image enhancement. For example, Wave-Mamba shows an LPIPS score of 0.185, while UHDformer achieves a PSNR of 27.113dB with only 0.34M parameters, showcasing impressive quantitative results.
In contrast, UHDRes achieves the highest PSNR while maintaining an extremely compact model size of just 400K learnable parameters. This is largely attributed to our proposed dual-domain decoupled spectral modulation framework, which effectively leverages spatial structures and frequency-domain cues, enabling high-quality restoration with fewer computational overhead.

\begin{table}[t]
\centering
\caption{Comparison of quantitative results on UHD-Haze. UHDRes demonstrates significant improvements across all metrics with a relatively small number of parameters.}
\label{tab:UHDHaze}
\setlength{\tabcolsep}{2pt}
\resizebox{\columnwidth}{!}{%

\begin{tabular}{cccccc}
\hline
\multirow{2}{*}{Method}          & \multirow{2}{*}{Source} & \multirow{2}{*}{Params$\downarrow$} & \multirow{2}{*}{PSNR$\uparrow$}   & \multirow{2}{*}{SSIM$\uparrow$}   & \multirow{2}{*}{LPIPS$\downarrow$} \\
                                 &                         &                                     &                                   &                                   &                                    \\ \hline
UHD~\cite{UHD}                   & ICCV'21                 & 34.5M                               & 18.043                            & 0.8113                            & -                                  \\
Restormer~\cite{Restormer}       & CVPR'22                 & 26.1M                               & 12.718                            & 0.6930                            & 0.456                              \\
Uformer~\cite{Uformer}           & CVPR'22                 & 20.6M                               & 19.828                            & 0.7374                            & 0.422                              \\
DehazeFormer~\cite{DehazeFormer} & TIP'23                  & 2.51M                               & 15.372                            & 0.7245                            & 0.400                              \\
TaylorFormer~\cite{TaylorFormer} & ICCV'23                 & 2.66M                               & 20.994                            & 0.9194                            & 0.312                              \\
UHDformer~\cite{UHDformer}       & AAAI'24                 & \textcolor{red}{\textbf{0.34M}}     & 22.586                            & 0.9427                            & 0.122                              \\
Histoformer~\cite{Histoformer}   & ECCV'24                 & 16.61M                              & 22.674                            & 0.9131                            & 0.189                              \\
MambaIR~\cite{MambaIR}           & ECCV'24                 & 2.01M                               & 21.453                            & 0.9054                            & 0.201                              \\
AdaIR~\cite{AdaIR}               & ICLR'25                 & 1.05M                               & 23.002                            & 0.9411                            & 0.124                              \\
D2Net~\cite{D2Net}               & WACV'25                 & 5.22M                               & 24.880                            & 0.9440                            & -                                  \\
UHDDIP~\cite{UHDDIP}             & TCSVT'25                & 0.81M                               & 24.699                            & \textcolor{blue}{\textbf{0.9520}} & \textcolor{blue}{\textbf{0.105}}   \\
ERR~\cite{ERR}                   & CVPR'25                 & 1.13M                               & \textcolor{blue}{\textbf{25.120}} & 0.9500                            & 0.119                              \\
UHDRes                           & Ours                    & \textcolor{blue}{\textbf{0.40M}}    & \textcolor{red}{\textbf{25.575}}  & \textcolor{red}{\textbf{0.9613}}  & \textcolor{red}{\textbf{0.101}}    \\ \hline
\end{tabular}

}
\end{table}

\begin{figure*}[t]
    \centering
    \includegraphics[width=\textwidth]{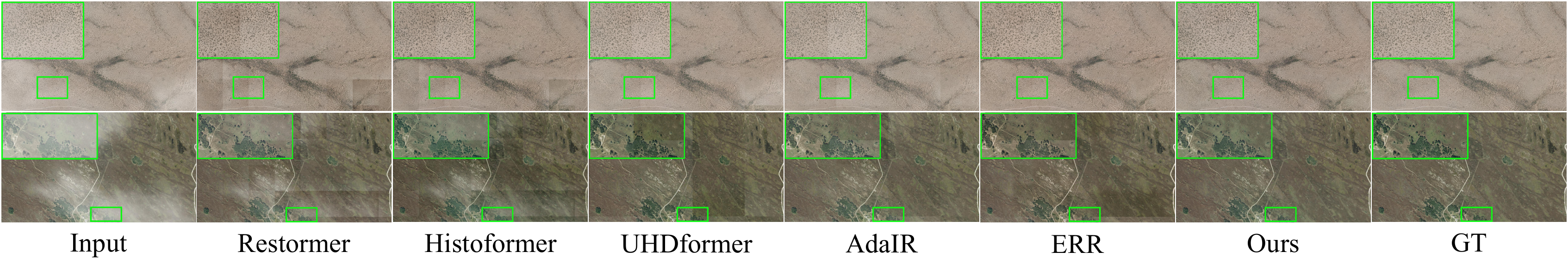}
    \caption{Visual comparison with other SOTA methods on the 8KDehaze-mini dataset. UHDRes yields more realistic results, whereas other methods exhibit noticeable boundaries.}
    \label{fig:8KDehaze}
\end{figure*}

\begin{figure*}[t]
    \centering
    \includegraphics[width=\textwidth]{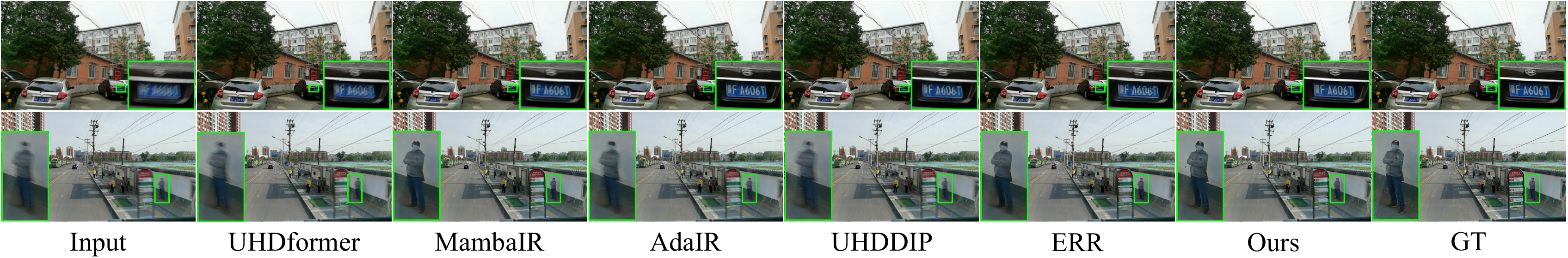}
    \caption{Visual comparison to SOTA methods on the UHD-Blur dataset. UHDRes restores clearer details and textures.}
    \label{fig:UHDBlur}
\end{figure*}

Fig.~\ref{fig:UHDLL} provides a visual comparison. The results show that under extremely low-light conditions, other methods such as UHDformer and AdaIR struggle to remove the heavy noise introduced by darkness, while methods like MambaIR tend to over-restore, resulting in distorted outputs.
Although ERR achieves relatively natural colors, its edges are not sufficiently clear. In contrast, our method achieves a more comprehensive and superior restoration effect.

\textbf{UHD image dehazing results.} 
Table~\ref{tab:UHDHaze} presents the quantitative results of various methods on the UHD-Haze dataset. UHDformer~\cite{UHDformer} achieves a PSNR of 22.586dB, while ERR~\cite{ERR} achieves a PSNR of 25.120dB and an SSIM of 0.9500, both of which demonstrate outstanding performance. However, these methods typically extract visual features from a low-resolution space, which inevitably leads to some loss of fine details.
In contrast, UHDRes achieves a PSNR of 25.575dB, an SSIM of 0.9613, and an LPIPS of 0.101, showing a significant advantage among all compared methods. This is largely due to our spatio-spectral fusion mechanism and the shared gated feed-forward network, which leverage the respective characteristics of the dual domains and fuse them effectively, thereby achieving a more comprehensive restoration.

From the qualitative results shown in Fig.~\ref{fig:UHDHaze}, it shows that in extremely challenging dense fog environments, where details of roads and buildings are severely lost, UHDformer and MambaIR, which do not model degradation information from a frequency domain perspective, fail to remove dense fog effectively. Owing to its insufficient modeling of high-frequency information, AdaIR may produce color distortions and blurry details. In contrast, our method achieves the best results in both dense fog removal and detailed texture restoration. This indicates that our decoupled spectral modulation block (DSMB), by leveraging decoupled learning tailored to the characteristics of frequency domain signals, achieves comprehensive and high-quality results.

\begin{table}[t]
\centering
\caption{
Comparison of quantitative results on the 8KDehaze-mini dataset. 
UHDRes achieves SOTA performance with a compact number of parameters.
}
\label{tab:8KDehaze}
\resizebox{\columnwidth}{!}{
\begin{tabular}{ccccc}
\hline
{Method}                    & {Source}               & {PSNR$\uparrow$} & {SSIM$\uparrow$} & {Params$\downarrow$} \\ \hline
{Restormer~\cite{Restormer}}     & {CVPR'22} & 24.901         & 0.9739         & 26.1M           \\
{Histoformer~\cite{Histoformer}}     & {ECCV'24} & 25.437         & 0.9689         & 16.61M           \\
{UHDformer~\cite{UHDformer}}     & {AAAI'24} & 27.590         & 0.9869         & \textcolor{red}{\textbf{0.34M}}           \\
{AdaIR~\cite{AdaIR}}         & {ICLR'25} & \textcolor{blue}{\textbf{29.506}}         & 0.9799         & 1.05M           \\
{ERR~\cite{ERR}}           & {CVPR'25} & 29.180         & \textcolor{blue}{\textbf{0.9878}}         & 1.13M           \\
UHDRes & Ours       & \textcolor{red}{\textbf{31.480}}         & \textcolor{red}{\textbf{0.9920}}         & \textcolor{blue}{\textbf{0.40M}}           \\ \hline
\end{tabular}
}
\end{table}

As shown in Table~\ref{tab:8KDehaze}, existing methods that perform well on the UHD-Haze dataset, such as UHDformer~\cite{UHDformer}, AdaIR~\cite{AdaIR}, and ERR~\cite{ERR}, experience performance degradation when processing 8K resolution images on the 8KDehaze-mini dataset. This is because high downsampling ratios and insufficient modeling of high-frequency components become more problematic with larger image sizes, resulting in suboptimal restoration quality.
In contrast, our spatio-spectral fusion module leverages multi-scale feature extraction and frequency-domain signal decoupling and restoration, enabling it to comprehensively learn both global degradation features and local detailed textures, thereby enhancing the restoration.

As shown in Fig.~\ref{fig:8KDehaze}, compared with other SOTA methods, our method achieves more natural restoration results, while other methods exhibit noticeable block boundaries. 
For UHD images, Vision Transformer-based methods, such as Restormer~\cite{Restormer}, Histoformer~\cite{Histoformer}, and AdaIR~\cite{AdaIR}, incur excessively high memory consumption. Even some methods with lightweight structural designs, like UHDformer~\cite{UHDformer} and ERR~\cite{ERR}, still face the challenge of rapidly increasing memory usage as the image resolution further increases, which also impacts their inference results. In contrast, our convolution and frequency-domain based method can more effectively reduce computational costs and enable direct full-resolution inference.

\begin{table}[t]
\centering
\caption{Comparison of quantitative results on UHD-Blur. UHDRes achieves SOTA performance with the smallest number of parameters.}
\label{tab:UHDBlur}
\setlength{\tabcolsep}{2pt}
\resizebox{\columnwidth}{!}{%
\begin{tabular}{cccccc}
\hline
\multirow{2}{*}{Method}        & \multirow{2}{*}{Source} & \multirow{2}{*}{Params$\downarrow$} & \multirow{2}{*}{PSNR$\uparrow$}   & \multirow{2}{*}{SSIM$\uparrow$}   & \multirow{2}{*}{LPIPS$\downarrow$} \\
                               &                         &                                     &                                   &                                   &                                    \\ \hline
MIMO-Unet++~\cite{MIMO-Unet++} & ICCV'22                 & 16.1M                               & 25.025                            & 0.7517                            & -                                  \\
Restormer~\cite{Restormer}     & CVPR'22                 & 26.1M                               & 25.210                            & 0.7522                            & 0.370                              \\
Uformer~\cite{Uformer}         & CVPR'22                 & 20.6M                               & 25.267                            & 0.7515                            & 0.385                              \\
Stripformer~\cite{Stripformer} & ECCV'22                 & 19.7M                               & 25.052                            & 0.7501                            & 0.374                              \\
FFTformer~\cite{FFTformer}     & CVPR'23                 & 16.6M                               & 25.409                            & 0.7571                            & 0.371                              \\
MixNet~\cite{MixNet}           & arXiv'24                & 5.22M                               & 29.430                            & 0.8550                            & -                                  \\
UHDformer~\cite{UHDformer}     & AAAI'24                 & \textcolor{blue}{\textbf{0.44M}}    & 28.821                            & 0.237                             & 0.237                              \\
MambaIR~\cite{MambaIR}         & ECCV'24                 & 2.01M                               & 30.400                            & 0.8694                            & 0.246                              \\
TriFormer~\cite{TriFormer}     & ICASSP'25               & 1.09M                                & 28.91                             & 0.8450                            & -                                  \\
AdaIR~\cite{AdaIR}             & ICLR'25                 & 1.05M                               & \textcolor{blue}{\textbf{30.838}} & \textcolor{blue}{\textbf{0.8816}} & \textcolor{blue}{\textbf{0.188}}   \\
D2Net~\cite{D2Net}             & WACV'25                 & 5.22M                               & 30.460                            & 0.8720                            & -                                  \\
UHDDIP~\cite{UHDDIP}           & TCSVT'25                & 0.81M                               & 29.517                            & 0.8585                            & 0.213                              \\
ERR~\cite{ERR}                 & CVPR'25                 & 1.13M                               & 29.720                            & 0.8610                            & 0.206                              \\
UHDRes                         & Ours                    & \textcolor{red}{\textbf{0.40M}}     & \textcolor{red}{\textbf{31.453}}  & \textcolor{red}{\textbf{0.8899}}  & \textcolor{red}{\textbf{0.176}}    \\ \hline
\end{tabular}
}
\end{table}

\begin{figure*}[t]
    \centering
    \includegraphics[width=\textwidth]{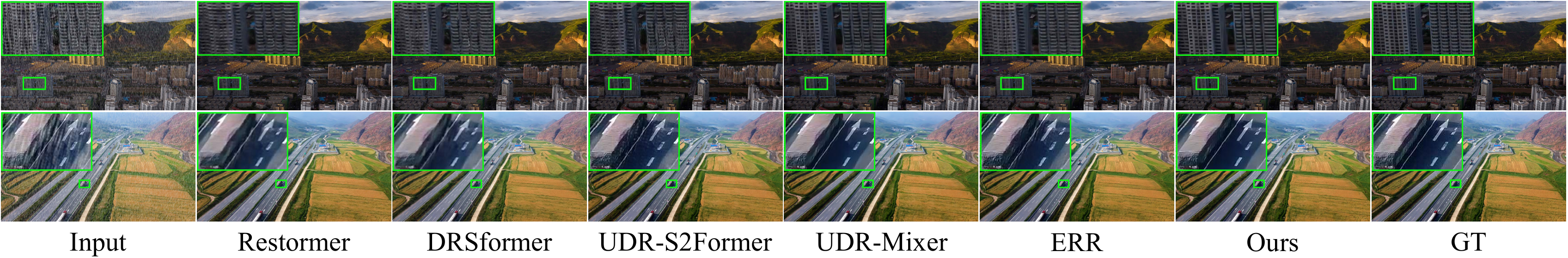}
    \caption{Visual comparison to SOTA methods on the 4K-Rain13k dataset. UHDRes achieves more refined rain streak removal.}
    \label{fig:Rain13k}
\end{figure*}

\begin{figure*}[!t]
    \centering
    \subfloat[]{\includegraphics[width=0.5\textwidth]{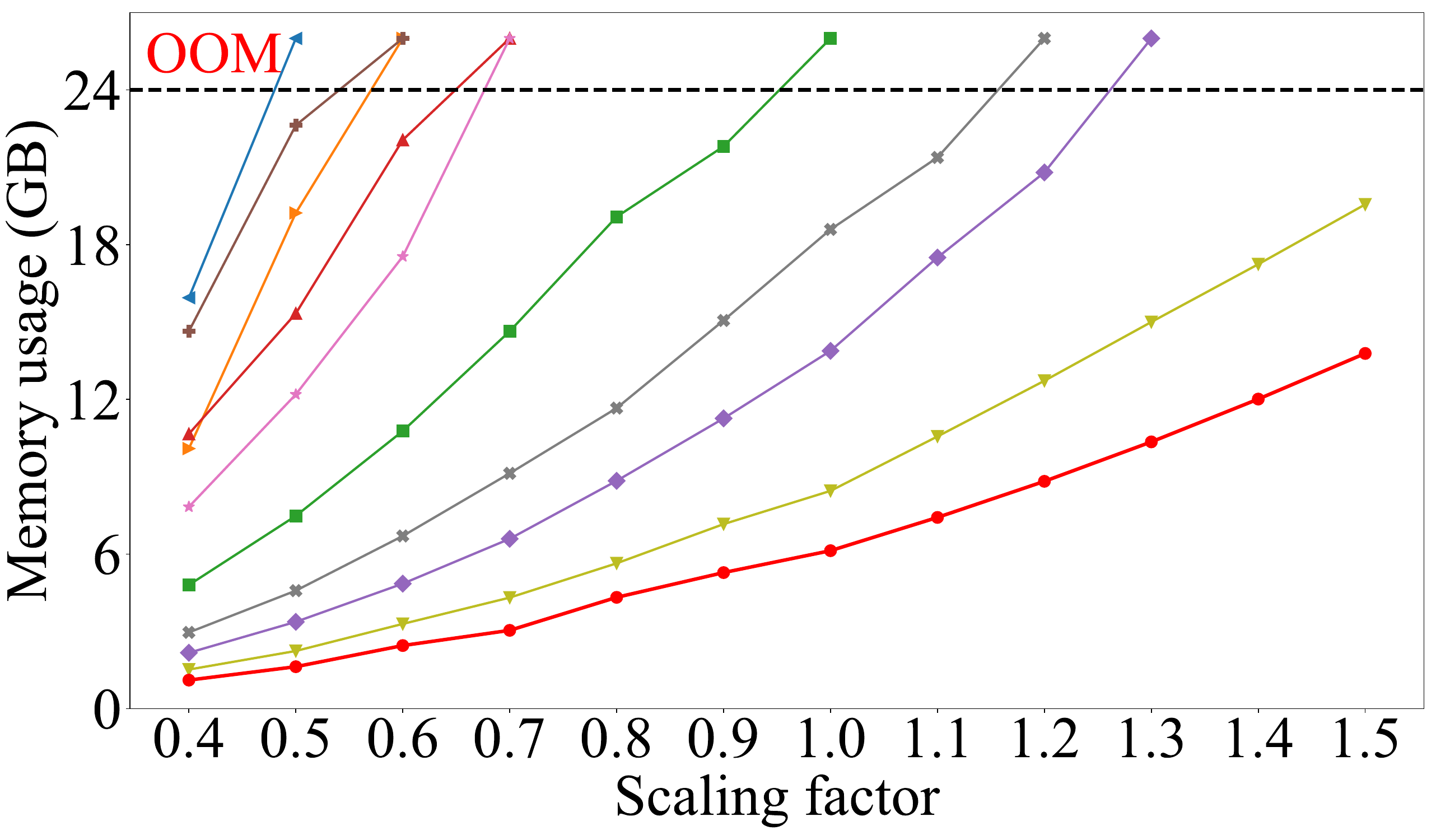}%
    \label{fig:memory}}
    \hfill
    \subfloat[]{\includegraphics[width=0.5\textwidth]{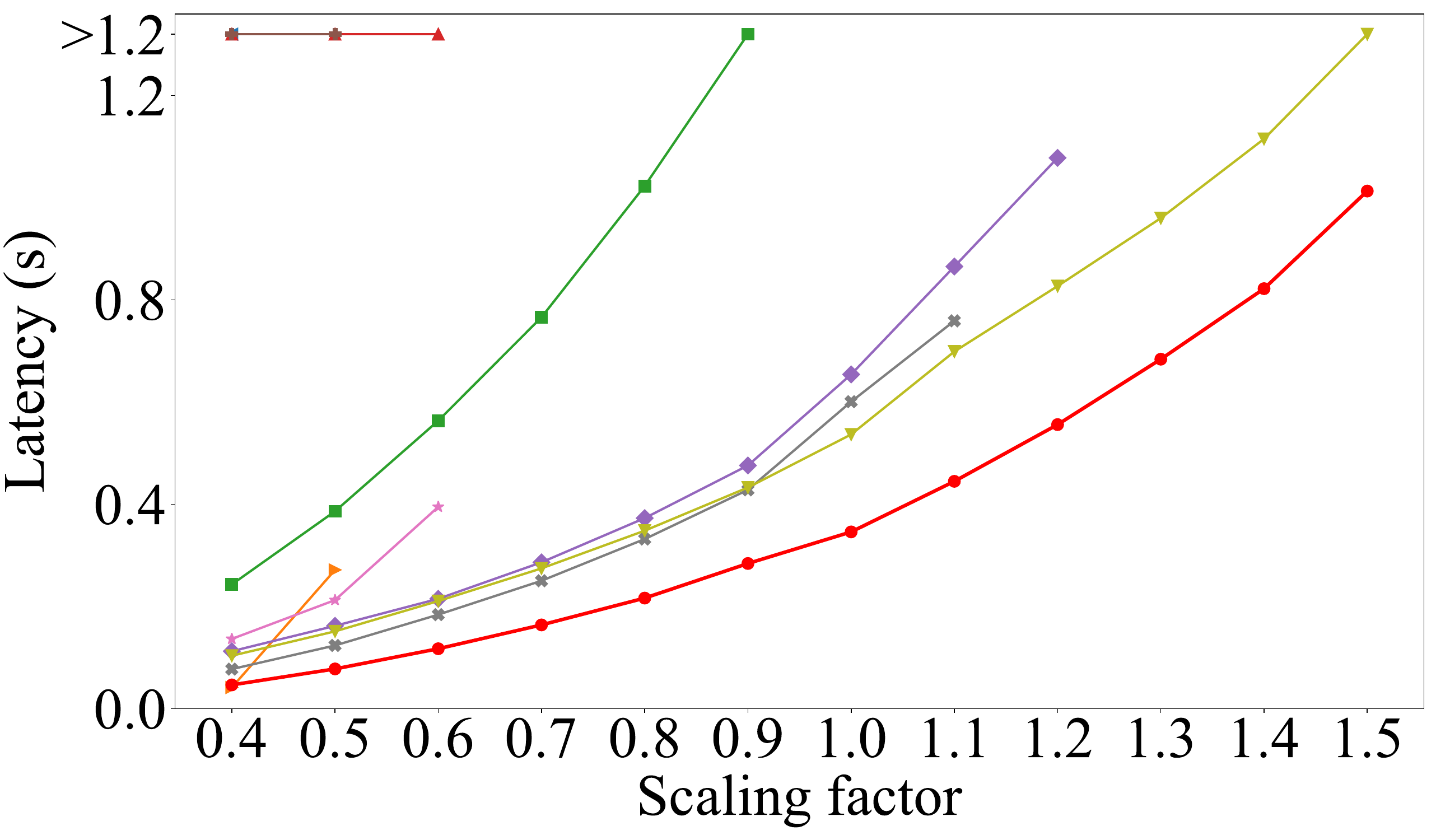}%
    \label{fig:latency}}
    
    \vspace{-2.6em}
 
    \subfloat[]{\includegraphics[width=0.95\textwidth]{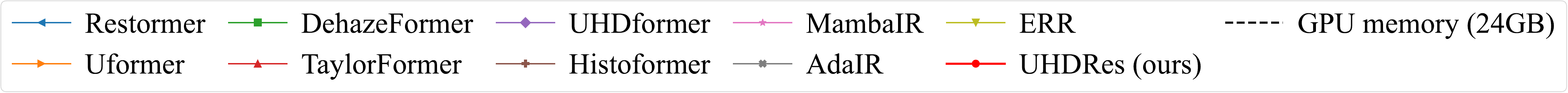}%
    \label{fig:legend}}
    \vspace{-1.5em}
    \caption{Peak GPU memory usage (left) and average inference latency (right) of compared methods at various scaling factors applied to a $2160 \times 3840$ resolution image.
    OOM indicates the out-of-memory issue. Our UHDRes exhibits the lowest GPU memory consumption and the fastest inference speed.
    }
    \label{fig:Performance}
\end{figure*}

\textbf{UHD image deblurring results.} 
Table~\ref{tab:UHDBlur} displays the quantitative results on the UHD-Blur dataset. Many methods demonstrate commendable performance. For example, AdaIR achieves a PSNR of 30.838dB and an LPIPS of 0.188.
However, the compared methods predominantly concentrate on low-frequency restoration, which mainly reflects the global degradation characteristics. In contrast, for 4K images with abundant structural details, local feature modeling plays an equally essential role.
Our method, UHDRes, achieves a PSNR of 31.453dB and significantly surpasses existing methods across quantitative results. This is largely credited to our decoupled spectral modulation block, which decouples and modulates the frequency-domain signal based on the characteristics of amplitude and phase. It effectively learns both global degradation features and detailed texture features, thereby achieving high-quality restoration.

From the qualitative results in Fig.~\ref{fig:UHDBlur}, it can be clearly observed that in the restoration results from recent image restoration methods, such as AdaIR and ERR, the license plates and pedestrians are relatively blurry. 
This indicates a failure to effectively restore the high-frequency details that are crucial for human visual perception. 
In contrast, UHDRes achieves a significantly clearer restoration. This further demonstrates that the design of our decoupled spectral modulation block enables our model not only to focus on global features but also to exhibit a powerful modeling capability for the abundant, detailed textures in UHD images.

\begin{table}[t]
\centering
\caption{Comparison of quantitative results on 4K-Rain13k. UHDRes achieves SOTA performance with the smallest number of parameters.}
\label{tab:Rain13k}
\setlength{\tabcolsep}{2pt}
\resizebox{\columnwidth}{!}{%
\begin{tabular}{cccccc}
\hline
\multirow{2}{*}{Method}       & \multirow{2}{*}{Source} & \multirow{2}{*}{Params$\downarrow$} & \multirow{2}{*}{PSNR$\uparrow$}   & \multirow{2}{*}{SSIM$\uparrow$}   & \multirow{2}{*}{LPIPS$\downarrow$} \\
                              &                         &                                     &                                   &                                   &                                    \\ \hline
JORDER-E~\cite{JORDER-E}      & TPRMI'19                & 4.21M                               & 30.460                            & 0.9117                            & 0.209                              \\
RCDNet~\cite{RCDNet}          & CVPR'20                 & 3.17M                               & 30.830                            & 0.9212                            & 0.196                              \\
SPDNet~\cite{SPDNet}          & ICCV'21                 & 3.04M                               & 31.810                            & 0.9223                            & 0.195                              \\
Restormer~\cite{Restormer}    & CVPR'22                 & 26.1M                               & 33.020                            & 0.9335                            & 0.173                              \\
IDT~\cite{IDT}                & TPRMI'22                & 16.41M                              & 32.910                            & 0.9479                            & 0.124                              \\
DRSformer~\cite{DRSformer}    & CVPR'23                 & 33.65M                              & 32.960                            & 0.9334                            & 0.171                              \\
UDR-S2Former~\cite{DRSformer} & ICCV'23                 & 8.53M                               & 33.360                            & 0.9458                            & \textcolor{blue}{\textbf{0.122}}   \\
UDR-Mixer~\cite{UDR-Mixer}    & arXiv'24                & 4.90M                               & 34.300                            & 0.9505                            & 0.133                              \\
ERR~\cite{ERR}                & CVPR'25                 & \textcolor{blue}{\textbf{1.13M}}                               & \textcolor{blue}{\textbf{34.479}} & \textcolor{blue}{\textbf{0.9516}} & \textcolor{red}{\textbf{0.120}}    \\
UHDRes                        & Ours                    & \textcolor{red}{\textbf{0.40M}}    & \textcolor{red}{\textbf{34.934}}  & \textcolor{red}{\textbf{0.9546}}  & \textcolor{blue}{\textbf{0.122}}   \\ \hline
\end{tabular}
}
\end{table}

\begin{figure*}[t]
    \centering
    \includegraphics[width=\textwidth]{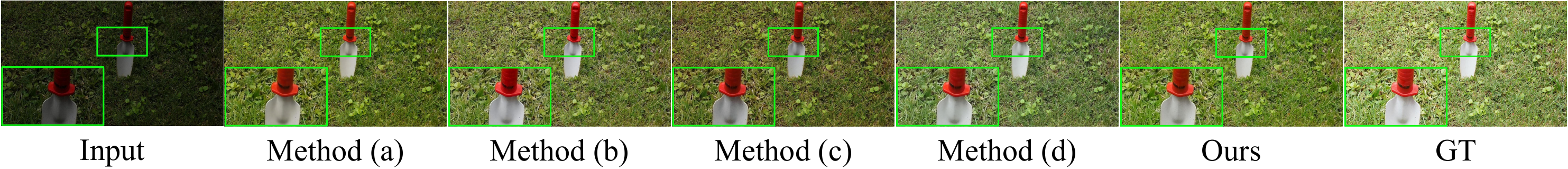}
    \caption{Ablation study on the proposed components. Method (a) is achieved by removing the MSCA, method (b) by removing the SAMU, method (c) by removing the SRU, and method (d) by removing the SGFN.}
    \label{fig:ablation_modules}
\end{figure*}

\textbf{UHD image deraining results.} 
Table~\ref{tab:Rain13k} shows the quantitative results on the 4K-Rain13k dataset. The SOTA method ERR achieves a PSNR of 34.479dB and an SSIM of 0.9516, demonstrating excellent performance. However, conventional deraining methods typically focus effectively on coarse-grained rain patterns but struggle to restore fine-grained rain streaks, resulting in partial blurring artifacts.
In contrast, UHDRes achieves 34.934dB PSNR and 0.9546 SSIM scores, significantly outperforming existing methods. This improvement mainly stems from the spectral amplitude modulation unit (SAMU), which effectively learns coarse-grained rain patterns in the frequency domain. In addition, the structural refinement unit (SRU) focuses on blurry regions affected by subtle rain streak interference, enabling comprehensive and high-quality restoration.

From the qualitative results in Fig.~\ref{fig:Rain13k}, we can observe that most rain streaks are effectively removed by methods such as UDR-S2Former, UDR-Mixer, and ERR.
However, degradation artifacts in buildings and vehicle shadows caused by fine rain streaks remain poorly restored. 
This indicates a failure to effectively capture fine-grained information crucial for comprehensive restoration. 
In contrast, UHDRes achieves significantly sharper and cleaner restoration results. This further demonstrates that our proposed spatio-spectral fusion mechanism enables our model not only to precisely identify degradation features but also to preserve critical original details, thereby achieving high-quality image restoration.

\textbf{Inference efficiency.}
UHD images possess ultra-high pixel characteristics, which is extremely important for model deployment in practical applications. This section primarily focuses on comparing the inference performance of the models. 

As depicted in Fig.~\ref{fig:Performance}, we visualize the maximum GPU memory usage (GB) and inference latency (s) for each model when inferring images of various resolutions. Specifically, images with a resolution of $2160 \times 3840$ were scaled to different degrees, and then each model performed inference on these scaled images to obtain the corresponding maximum GPU memory usage and inference latency. Note that these experiments were conducted on a single NVIDIA 4090 GPU. From Fig.~\ref{fig:Performance}, it can be observed that Restormer, TaylorFormer, and Histoformer exhibit high GPU memory usage and high latency during inference. Among them, TaylorFormer's latency reached 7.64 seconds when inferring at a resolution of $1290 \times 2304$, making it difficult to deploy in edge device scenarios. MambaIR and Uformer also show excessively high GPU memory usage during inference. Dehazeformer encounters CUDA out of memory when performing 4K image inference, thus preventing full-resolution inference. UHDformer and AdaIR also demonstrate a rapid increase in both memory usage and latency as image size increases. In contrast, UHDRes maintains the lowest GPU memory usage and minimal inference latency across all resolutions.

\begin{table}[t]
\centering
\caption{
Ablation study on the proposed components of UHDRes.
\ding{55} denotes removal, while \ding{51} indicates inclusion.
}
\label{tab:ablation_module}
\resizebox{\columnwidth}{!}{%
\begin{tabular}{ccccccc}
\hline
Method & MSCA   & SAMU   & SRU    & \multicolumn{1}{l}{SGFN} & PSNR$\uparrow$ & SSIM$\uparrow$ \\ \hline
(a)      & \ding{55} & \ding{51} & \ding{51} & \ding{51}                   & 27.14          & 0.9274         \\
(b)      & \ding{51} & \ding{55} & \ding{51} & \ding{51}                   & 26.83          & \textcolor{red}{\textbf{0.9311}}         \\
(c)      & \ding{51} & \ding{51} & \ding{55} & \ding{51}                   & \textcolor{blue}{\textbf{27.34}}          & 0.9272         \\
(d)      & \ding{51} & \ding{51} & \ding{51} & \ding{55}                   & 27.08          & \textcolor{blue}{\textbf{0.9300}}         \\
Ours   & \ding{51} & \ding{51} & \ding{51} & \ding{51}                   & \textcolor{red}{\textbf{27.69}}          & \textcolor{red}{\textbf{0.9311}}         \\ \hline
\end{tabular}
}
\end{table}

\subsection{Ablation study}
To demonstrate the effectiveness of each proposed component and the sensitivity of its parameters, we conducted comprehensive ablation experiments. All ablation studies are conducted on the UHD-LL dataset.

\textbf{Ablation study of the proposed components.}
Table~\ref{tab:ablation_module} presents the ablation results of each proposed module. We demonstrate their effectiveness by removing the corresponding modules. (a) When the multi-scale context aggregator (MSCA) was removed, the PSNR decreased by 0.55dB. This is primarily due to the lack of multi-scale features extracted by MSCA in the spatial domain, which, to some extent, diminishes the model's decoupling capability and prevents effective restoration. 
(b) When the spectral amplitude modulation unit (SAMU) is removed, the PSNR drops by as much as 0.86dB. This is mainly because SAMU plays an explicit modeling role in frequency domain information modulation, and its absence significantly impedes the model's restoration.
(c) When the structural refinement unit (SRU) was removed, the SSIM only reached 0.9272. Since SRU is designed to compensate for high-frequency phase in the spatial domain, it effectively restores fine-grained textures, thereby enhancing the perceptual quality of the restoration results. (d) When the shared gated feed-forward network (SGFN) is removed, the PSNR decreases by 0.61dB. This is primarily due to the lack of SGFN's effective fusion of features modulated in dual domains, which diminishes the feature representation learning capability and consequently leads to poorer quality restoration results.

Fig.~\ref{fig:ablation_modules} shows the qualitative results of UHDRes and the results after separately removing each module. It's clear that compared to UHDRes, removing any single component leads to poor restoration quality, specifically with color deviations and darker brightness. This demonstrates the validity of our proposed dual-domain decoupled spectral modulation framework for UHD image restoration, which exploits spatial structures and frequency-domain cues in a decoupled manner, thereby achieving higher restoration quality.

\begin{table}[t]
\centering
\caption{
Ablation study of MSCA with various sizes of convolutional kernels on the UHD-LL dataset. The ``\textit{id}'' indicates the identity mapping operation. 
}
\label{tab:ablation_LKS}
\resizebox{\columnwidth}{!}{%

\begin{tabular}{ccccc}
\hline
Method                             & \makebox[0.12\textwidth][c]{Kernel sizes}                        & PSNR$\uparrow$   & SSIM$\uparrow$   & Params$\downarrow$ \\ \hline
{(e)}             & {$[id,3,7,11]$}  & 26.866 & 0.9281 & \textcolor{red}{\textbf{382.79K}}   \\
{(f)}             & {$[id,3,9,15]$}  & \textcolor{blue}{\textbf{27.383}} & 0.9306 & 408.90K   \\
{(g)}             & {$[id,5,11,17]$} & 27.093 & \textcolor{red}{\textbf{0.9328}} & 431.94K   \\
{(h)}             & {$[id,7,11,15]$} & 27.230 & 0.9295 & 424.26K   \\
{(i)}             & {$[id,7,13,19]$} & 27.335 & 0.9271 & 459.59K   \\
{{Ours}} & {$[id,5,9,13]$}  & \textcolor{red}{\textbf{27.693}} & \textcolor{blue}{\textbf{0.9311}} & \textcolor{blue}{\textbf{401.22K}}   \\ \hline
\end{tabular}
}
\end{table}

\textbf{Ablation study of multi-scale context aggregator (MSCA).}
Table~\ref{tab:ablation_LKS} details the impact of different convolutional kernel size combinations on the model performance within the MSCA module. In the spatial domain, we adopted a parallel four-branch structure, where one branch performs identity mapping and the other three branches execute large kernel depthwise convolutions of different sizes.

The high pixel density of UHD images necessitates a larger receptive field for the model. Large kernel convolutions, owing to their larger receptive fields, can effectively model long-range dependencies. However, this does not imply that larger kernel sizes are always superior. Excessively large convolution kernels not only lead to a significant increase in model parameters but also make the network's parameter optimization challenging. From the experimental results, it can be clearly observed that the combination utilizing identity mapping and convolution kernels with sizes of $5 \times 5$, $9 \times 9$, and $13 \times 13$ achieves the best performance. 

\begin{table}[t]
\centering
\caption{
Ablation study of SAMU on the UHD-LL dataset. \ding{55} denotes the absence of modeling, while \ding{51} represents the inclusion of modeling.
}
\label{tab:ablation_SAMU}
\resizebox{\columnwidth}{!}{%
\begin{tabular}{ccccc}
\hline
\makebox[0.06\textwidth][c]{Method} & \makebox[0.06\textwidth][c]{Amplitude}              & \makebox[0.06\textwidth][c]{Phase}                       & \makebox[0.06\textwidth][c]{PSNR$\uparrow$}                                     & \makebox[0.06\textwidth][c]{SSIM$\uparrow$}                                                                                                      \\ \hline
(j)                       & \ding{55} & \ding{51} & 27.48                                                               & 0.9297                                                               \\
(k)                       & \ding{51} & \ding{51} & 27.10                                                               & 0.9297                                                               \\
Ours                    & \ding{51} & \ding{55} & \textcolor{red}{\textbf{27.69}} & \textcolor{red}{\textbf{0.9311}} \\ \hline
\end{tabular}
}
\end{table}

\begin{figure*}[t]
    \includegraphics[width=\textwidth]{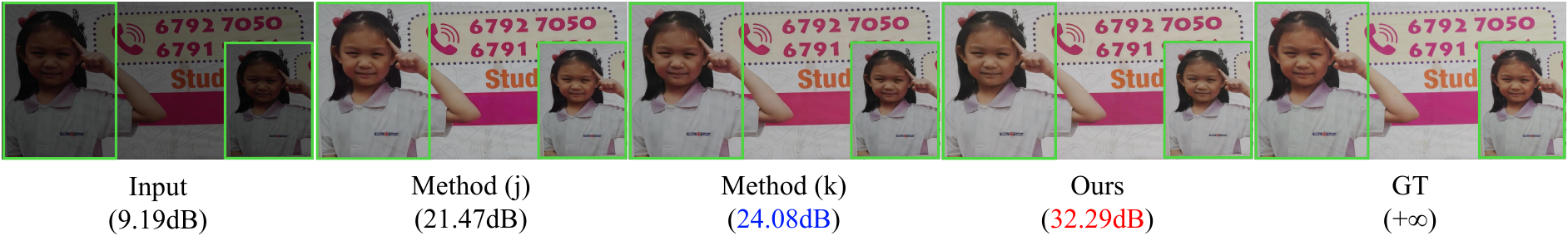} 
    \caption{
    Ablation study of selective modeling in SAMU. 
    The method (j) denotes no amplitude spectrum modulation in SAMU, and method (k) denotes simultaneous modulation of both amplitude and phase spectra in SAMU. Our simplified modulation of the amplitude spectrum yielded the best overall restoration.
    }
    \label{fig:ablation_1}
\end{figure*}

\begin{figure*}[t]
    \centering
    \includegraphics[width=\textwidth]{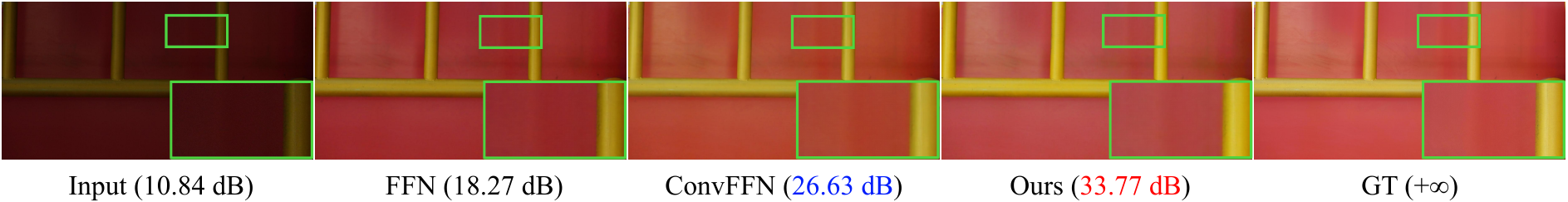} 
    \caption{Ablation study of feed-forward networks. Our SGFN restored more realistic colors compared to other FNNs.}
    \label{fig:ablation_ffn}
\end{figure*}

\textbf{Ablation study of spectral amplitude modulation unit (SAMU).}
Based on the observations shown in Fig.~\ref{fig:sec2}, we designed relevant ablation experiments to verify the effectiveness of the spectral amplitude modulation unit (SAMU).
As shown in Table~\ref{tab:ablation_SAMU}, we conducted experiments on three sets of comparative experiments: (j) modulating only the phase spectrum in SAMU, (k) simultaneously modulating both amplitude and phase spectra in SAMU, and (Ours) modulating only the amplitude spectrum in SAMU. The experimental results indicate that selectively modulating only the amplitude spectrum in SAMU, while implicitly modeling the high-frequency phase spectrum using our structural refinement unit (SRU), achieved the best results. 
These results confirm the value of frequency-domain modeling. However, phase information is highly sensitive, and minor prediction errors may cause large performance drops. By implicitly modeling the phase via identity mapping, we allow the model to concentrate on learning the amplitude spectrum, while our SRU implicitly refines structural details.

Fig.~\ref{fig:ablation_1} presents the qualitative comparison among method (j), method (k), and our full model. Notably, modeling only the amplitude spectrum in the frequency domain achieves a PSNR improvement of 8.21dB over jointly modeling both amplitude and phase, and more effectively suppresses darkness in the image. We attribute this improvement to the inherent coupling between amplitude and phase spectra. Simultaneously optimizing both introduces additional complexity, often leading to unstable training dynamics and forcing the model into a ``poor compromise'' that limits restoration quality.

\begin{table}[t]
\centering
\caption{
Ablation study of feed-forward networks on the UHD-LL dataset. We replaced the FFN part and adjusted the channels to maintain a similar number of parameters.
}
\label{tab:ablation_FFN}
\resizebox{\columnwidth}{!}{%

\begin{tabular}{ccccc}
\hline
Method              & FFN     & ConvFFN & GatedFFN & SGFN (ours)            \\ \hline
PSNR$\uparrow$                & 26.814  & \textcolor{blue}{\textbf{26.986}}  & 26.894   & \textcolor{red}{\textbf{27.693}} \\
SSIM$\uparrow$                & 0.9309  & \textcolor{red}{\textbf{0.9321}}  & 0.9310   & \textcolor{blue}{\textbf{0.9311}}          \\
Param$\downarrow$            & 404.87K & 420.23K & \textcolor{red}{\textbf{386.06K}}  & \textcolor{blue}{\textbf{401.22K}}  \\ \hline
\end{tabular}
}
\end{table}

\begin{figure*}[t]
    \centering
    \includegraphics[width=\textwidth]{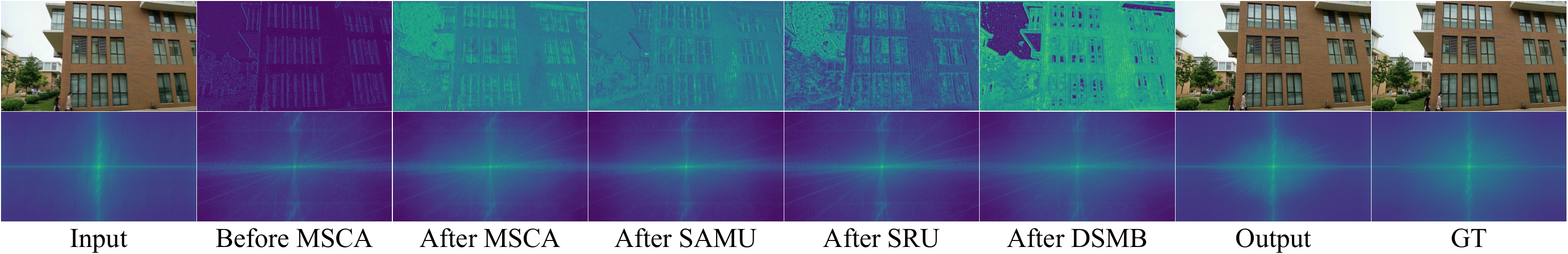} 
    \caption{Visual effects of SSFM with image/feature domain (top) and amplitude spectra (bottom) on the UHD-Blur dataset. In Fig.~\ref{fig:Model} (a) (UHDRes), we visualize the SSFM of the second DAEB in its bottleneck layer.
        }
    \label{fig:CFAM_feature}
\end{figure*}

\textbf{Ablation study of feed-forward networks.}
As presented in Table~\ref{tab:ablation_FFN}, ablation studies evaluating our shared gated feed-forward network (SGFN) against established feed-forward network (FFN) variants, including FFN~\cite{ffn}, ConvFFN~\cite{RepLKNet}, and GatedFFN~\cite{gatedffn}, consistently demonstrate SGFN's superior performance. 

Fig.~\ref{fig:ablation_ffn} further illustrates that SGFN yields color restoration more consistent with the ground-truth image. This effectiveness is attributed to the SGFN's weight-sharing mechanism. It facilitates learning an enhanced feature representation across both spatial and frequency domains. Additionally, the horizontally and vertically striped convolutional gating design enhances feature fusion. As a consequence, These operations collectively contribute to higher-quality restoration.

\subsection{More analysis of visualization.} 
Fig.~\ref{fig:CFAM_feature} illustrates the effect of each component within the spatio-spectral fusion module (SSFM) on feature representations.
Before the multi-scale context aggregator (MSCA), the input features exhibit low contrast, and their spectral responses retain degradation patterns similar to the original image.
After passing through the MSCA, the intermediate features demonstrate significantly enhanced contrast and richer details. 
Following the spectral amplitude modulation unit (SAMU), which models amplitude spectra where most energy resides in low-frequency components, the structural characteristics become more pronounced.
The structural refinement unit (SRU), primarily composed of local convolutional operations, implicitly compensates for phase spectral information, producing finer texture details. 
After the dual-domain adaptive enhancement block (DSMB), fusion across spatial and frequency domains further refines both structure and texture, while effectively suppressing residual degradation patterns.
These observations collectively verify the capability of SSFM to achieve comprehensive restoration in both spatial and frequency domains.

\begin{figure}[t]
    \centering
    \includegraphics[width=\columnwidth]{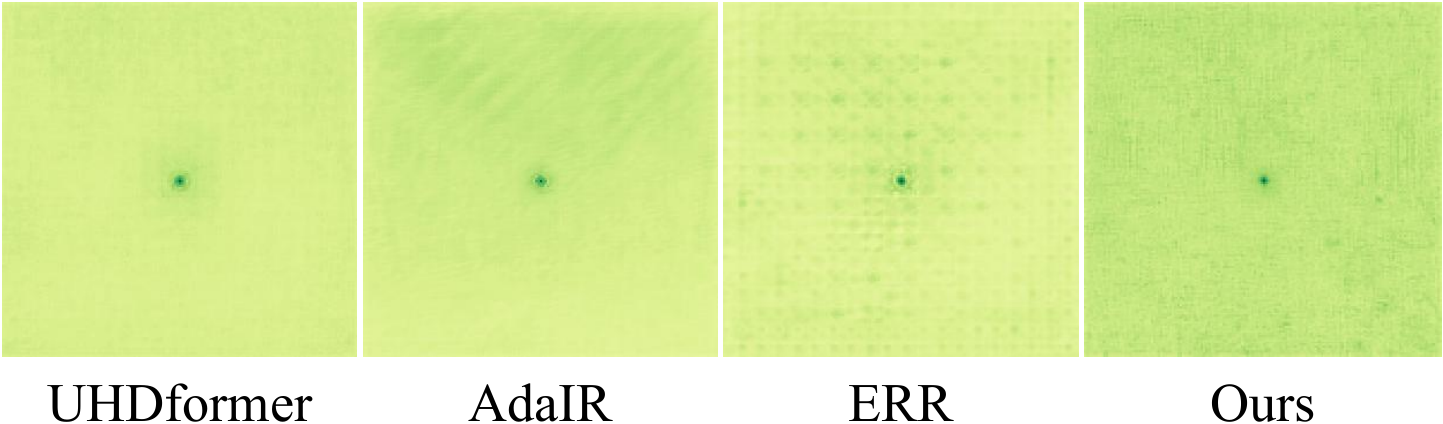} 
    \caption{Visualization of the effective receptive field for UHDformer~\cite{UHDformer}, AdaIR~\cite{AdaIR}, ERR~\cite{ERR} and our UHDRes.}
    \label{fig:ERF}
\end{figure}

As shown in Fig.~\ref{fig:ERF}, we also visualize effective receptive fields for vision transformer-based methods~\cite{UHDformer,AdaIR}, hybrid architecture methods~\cite{ERR}, and our method. Darker shades indicate larger effective receptive fields, reflecting the model's modeling capability. It can be observed that AdaIR exhibits ripple-like triangular patterns in the upper region, while ERR displays radially decreasing dot patterns around its center point. These characteristics reveal directional biases in both methods. Although UHDformer shows relatively uniform coverage, its shades remain comparatively lighter. In contrast, our method demonstrates more uniform and larger effective receptive fields, thus indicating that UHDRes possesses effective global modeling capability.

\section{Conclusion}
We have presented {UHDRes}, a novel dual-domain decoupled spectral modulation framework for UHD image restoration. 
By explicitly modeling the robust amplitude spectrum in the frequency domain while implicitly restoring phase through subsequent spatial-domain refinement, our UHDRes achieves an effective balance between restoration quality and computational efficiency. 
The proposed spatio-spectral fusion mechanism and shared gated feed-forward mechanisms are integrated into dual-domain adaptive enhancement blocks, enabling the model to effectively handle diverse degradation patterns while guiding features to an enhanced representation and maintaining low overhead.
Extensive experiments across multiple UHD image restoration tasks have demonstrated that our UHDRes not only delivers the SOTA restoration performance with only 400K parameters but also significantly improves runtime efficiency. 

\bibliographystyle{IEEEtran}
\bibliography{ref_v2}

\end{document}